# The Making of Delta Sunspots

Ronald L. Moore[1,2], Sanjiv K. Tiwari[3,4], V. Aparna[3,4], Navdeep K. Panesar[3,5], and Alphonse C. Sterling[2]

[1]Center for Space Plasma and Aeronomic Research (CSPAR), UAH, Huntsville AL, 35805 USA;
[2]NASA Marshall Space Flight Center, Huntsville, AL 35812, USA
ronald.l.moore@nasa.gov
[3]Lockheed Martin Solar Astrophysics Laboratory, 3251 Hanover Street Building 203, Palo Alto, CA 94306, USA
[4]Bay Area Environmental Research Institute, NASA Research Park, Moffett Field, CA 94035, USA; tiwari@lmsal.com
[5]SETI Institute, 339 Bernardo Ave, Mountain View, CA 94043, USA

## Abstract

We explore what fraction of delta sunspots in which the polarity inversion line (PIL) is sharp in photospheric magnetograms are made from a writhe kink in an emerging twisted flux rope. We searched simultaneous full-disk magnetograms and continuum images from Helioseismic and Magnetic Imager (HMI) on Solar Dynamics Observatory (SDO) to find 28 random sharp-PIL delta sunspots that are born well on the disk. Only one of these is made from a single newly emerged bipolar magnetic region (BMR) and therefore is a candidate for being made from a single emerging writhe-kinked flux rope. That outcome indicates that few, if any, sharp-PIL delta sunspots are made by a single emerging writhe-kinked flux rope. That is the main new finding of this paper. Each of the other 27 is made by merging of two or more emerging or emerged BMRs. We name delta-sunspot genesis from a single BMR Type I genesis. We identify another three genesis types among the other 27 delta sunspots: Type II, Type III, and Type IV. We present an observed example genesis for each of the four genesis types, and for each example present schematic drawings depicting our scenario(s) for the cause of that example genesis. The core idea of each scenario is that the delta sunspot is made by packing opposite-polarity magnetic flux together by advection into a convection downflow.

*Unified Astronomy Thesaurus concepts:* Delta sunspots (1979); Solar active region magnetic fields (1975); Solar active region velocity fields (1974); Solar convective zone (1998)

## 1. Introduction

The magnetic field filling the Sun's chromosphere and corona evidently all comes from magnetic flux-rope $\Omega$ loops that bubble up from below the photosphere (e.g., C. Zwaan, 1978; R. Ishikawa, S. Tsuneta, & J. Jurcak 2010; L. van Driel-Gesztelyi & L. M. Green 2015; Y. Fan 2021). A lone recently emerged flux-rope $\Omega$ loop, i.e., one that emerges well away from other comparably large or larger emerged or emerging $\Omega$ loops, is seen as a lone bipolar magnetic region (BMR) in photospheric magnetograms (e.g., L. van Driel-Gesztelyi & L. M. Green 2015; R. L. Moore, S. K. Tiwari, N. K. Panesar, & A. C. Sterling 2020). Usually, the BMR roughly fills an elliptical area having its major diameter roughly twice its minor diameter, and the BMR's opposite-polarity flux is concentrated in and roughly fills opposite long halves of the ellipse (e.g., R. L. Moore, S. K. Tiwari, N. K. Panesar, & A. C. Sterling 2020). If the lone flux-rope $\Omega$ loop has magnetic flux $\Phi <\sim 10^{20}$ Mx, the BMR has no sunspots or sunspot pores; for $10^{20} <\sim \Phi <\sim 5 \times 10^{21}$ Mx, the BMR has sunspot pores but no full-fledged sunspots with penumbra; for $\Phi >\sim 5 \times 10^{21}$ Mx, the BMR has full-fledged sunspots (L. van Diel-Gesztelyi & L. M. Green 2015). For the Sun's largest lone BMRs, $\Phi \sim 3 \times 10^{22}$ Mx, and each of the BMR's two opposite-polarity flux domains has one or more large sunspots (Y. M. Wang & N. R. Sheeley 1989; R. L. Moore, S. K. Tiwari, N. K. Panesar, & A. C. Sterling 2020). Each BMR that has one or more full-fledged sunspots – as well as each BMR that has sunspot pores but no full-fledged

sunspots – is called a solar active region (AR) and is given a chronological AR number by NOAA (National Oceanic and Atmospheric Administration).

Solar eruptions of all sizes – from the greatest flares and coronal mass ejections (CMEs) down through subflares and coronal jets, and on down through campfires, nanoflares, and picoflares – are powered by release of energy from the magnetic field in which the eruption occurs, and the release is enabled and sustained by reconnection of that field (e.g., T. Hirayama 1974; Z. Svestka 1976; P. A. Sturrock 1980; H. Zirin 1988; R. L. Moore & G. Roumeliotis 1992; K. Shibata, 1998; R. L. Moore, A. C. Sterling, H. Hudson, & J. R. Lemen 2001; R. L. Moore, A. C. Sterling, & D. A. Falconer 2015; K. Shibata & T. Magara 2011; S. K. Tiwari, C. E. Alexander, A. R. Winebarger, & R. L. Moore 2014; A. C. Sterling, R. L. Moore, D. A. Falconer, & M. Adams 2015; N. K. Panesar, A. C. Sterling, R. L. Moore, et al. 2018, 2019; S. Toriumi & H. Wang 2019; N. Panesar, S. K. Tiwari, D. Berghmans, et al. 2021; S. K. Tiwari, N. K., Panesar, R. L. Moore, et al. 2019; S. K. Tiwari, V. H. Hansteen, B. De Pontieu, N. K. Panesar, & D. Berghmans 2022; N. E. Raouafi, G. Stenborg, D. B. Seaton, et al 2023; L. P. Chitta, A. N. Zhudov, D. Berghmans, et al. 2023).

In this paper, the term "major flare eruption" or "major flare" refers to any solar eruption that produces a thermal X-ray burst that peaks at or above the M 1.0 level in GOES 1-8 Å thermal X-ray flux ($\geq 10^{-2}$ erg cm$^{-2}$ s$^{-1}$), i.e., any flare eruption that produces a GOES class M or X thermal X-ray burst. In flares that produce a thermal X-ray burst detectable by GOES, the thermal X-ray plasma fills one or more arcades of coronal magnetic loops, each arcade envelops an underlying polarity inversion line (PIL) in photospheric magnetic flux, and the thermal X-ray plasma results from reconnection of the magnetic field that holds it (e.g., M. E. Machado, R. L. Moore, A.M. Hernandez, et al. 1988).

A delta sunspot is a sunspot that has positive-polarity umbra and negative-polarity umbra close together within surrounding penumbra (e.g., M. J. Martres & A. Bruzek 1977). Usually, the PIL between a delta sunspot's two opposite-polarity magnetic flux domains is sharp in photospheric magnetograms (e.g., R. L. Moore, D. A. Falconer, & A. C. Sterling 2012; G. Chintzoglou, J. Zhang, M. C. M. Cheung, & M. Kazachenko 2019). Major flares often occur in magnetic field enveloping the PIL of a delta sunspot (H. Zirin & M. A. Liggett 1987; H. Wang, F. Tang, & H. Zirin 1991; F. Tang & H. Wang, 1993; T. T. Ishii, H. Kurokawa, T. T. Takeuchi 1998; M. Kubo, T. Yokoyama, Y. Katsukawa, et al. 2007; G. Chintzoglou, J. Zhang, M. C. M. Cheung, & M. Kazachenko 2019; S. Toriumi & H. Wang 2019). Hence, an obvious question in connection with the production of major flares is: How are delta sunspots made?

It is observed that some delta sunspots are made by the emergence of a flux-rope Ω loop that has one end near the opposite-polarity end of another emerging or emerged flux-rope Ω loop, each of which Ω loops is large enough to make a BMR that has a full-fledged sunspot in each of its two opposite-polarity flux domains. As the emergence continues, the positive end of one BMR becomes merged with the negative end of the other BMR, thereby making a delta sunspot. For examples of observed delta sunspots being made this way, see H. Zirin & M. A. Liggett (1987), G. Chintzoglou, J. Zhang, M. C. M Cheung, & M. Kazachenko (2019), and S. Toriumi & H. Wang 2019.

Another possibility for making delta sunspots is the following. It is imagined that sometimes, before a rising flux-rope Ω loop begins emerging through the photosphere, the magnetic field in the Ω loop's flux rope is greatly enough twisted about the flux-rope's center line that the top of the Ω loop has been given enough writhe to greatly change the horizontal direction of the top of the emerging Ω loop from the horizontal direction it would have had if the flux rope's field were not twisted enough for substantial writhe to have occurred. It has been further proposed that a delta sunspot can be formed by the emergence of a greatly writhed (i.e., kinked) top of a single emerging flux-rope Ω loop (K. Tanaka 1991). MHD simulation of the emergence of a writhe-kinked flux rope has demonstrated that this way of making delta sunspots is a physically viable possibility (Y. Fan, G. E. Zweibel, M. G. Linton, & G. Fisher 1999).

The present paper explores what fraction of the population of delta sunspots in which the PIL is sharp comes from merger of opposite-polarity feet of two or more emerging or emerged flux-rope Ω loops, and what fraction comes from merger of the two feet of a single flux-rope Ω loop. For that, we found 28 sharp-PIL delta sunspots for each of which the genesis is observed. Of those, 27 are made by merger of opposite-polarity feet of two or more emerging or emerged Ω loops; only one is made from a single Ω loop. For that one, we depict two possibilities: (1) the top of the emerging Ω loop is not writhe-kinked; (2) the top of the emerging Ω loop is writhe-kinked. We conclude that few sharp-PIL delta sunspots (no more than several percent) are made from a single flux-rope Ω loop, either not-kinked or kinked, and for any of those few, the top of the emerging flux-rope Ω loop might not be writhe-kinked, contrary to the model of K. Tanaka (1991) and Y. Fan, E. G. Zweibel, M. G. Linton, & G. H. Fisher (1999).

## 2. Data and Methods

The data are full-disk line-of-sight magnetograms and full-disk continuum images from Helioseismic and Magnetic Imager (HMI; P. H. Scherrer, J. Schou, R. I Bush, et al. 2012) of Solar Dynamics Observatory (SDO; W. D. Pesnell, B. J. Thompson, & P. C. Chamberlin 2012). These have 0.5" pixels and 45 s cadence.

Using Helioviewer (https://helioviewer.org), we searched through over ten years of HMI full-disk magnetograms and HMI full-disk continuum images (from 2013 November through 2024 June) for what we call in-sunspot sharp PILs. We define an in-sunspot sharp PIL to be any PIL that is sharp (<~ 3 pixels wide) in Helioviewer HMI magnetograms and that Helioviewer HMI continuum images show to be in a sunspot. By this definition, an in-sunspot sharp PIL can of course be a delta sunspot's PIL, but some are not. An in-sunspot sharp PIL that is not the PIL of a delta sunspot is one for which one of the two opposite-polarity flux domains sandwiching the PIL has penumbra but no umbra, instead of having both penumbra and umbra as in a delta sunspot.

We selected for study of the birth of the in-sunspot sharp PIL those in-sunspot sharp PILs for which the on-disk evolution of the sharp PIL's active region (AR) can be tracked back through the birth of the sharp PIL in the HMI magnetograms and continuum images. Our search yielded 29 such in-sunspot sharp PILs. They are listed in Table 1. Other than each being required to have an on-disk birth, these 29 in-sunspot sharp PILs are a random sample: each happened at random, by chance. Each but one of the 29 PILs in Table 1 is the PIL of a delta sunspot, i.e., has umbra as well as penumbra on both sides. Only one of the PILs in Table 1 (number 13) has only penumbra on one side.

We used Solar Monitor (https://www.solarmonitor.org) to get the NOAA AR number of each active region in which a selected in-sunspot sharp PIL is born. An AR's NOAA number is the chronological number it is given by the National Oceanic and Atmospheric Administration (NOAA).

For two example in-sunspot sharp PILs, we use coronal EUV images from SDO's Atmospheric Imaging Assembly (AIA; J. R. Lemen, A. M. Title, D. J. Akin, et al. 2012) to show for one PIL the direction of shear of the magnetic field enveloping the PIL, and to show for the other PIL that the overall form of the coronal field of the PIL's active region is S-shape sigmoidal. AIA's coronal EUV images have 0.6" pixels and 12 s cadence.

## 3. Results

### 3.1. Four Types of In-sunspot Sharp PIL Genesis

In our set of 29 in-sunspot sharp PILs, the PILs are made in four classes or types of ways, which we call Type I, Type II, Type III, and Type IV. In Figure1, each of these is defined with words and depicted by a schematic. In these schematics, the BMRs are viewed from above, with solar north up and solar west right; positive flux is red and negative flux is blue. In each schematic, the horizontal arrow represents the time interval of evolution between early in the emergence of one or two BMRs to the left of the arrow and later, to the right of the arrow, when further emergence of the BMR(s) has made an in-sunspot sharp PIL.

The schematic for Type I genesis in Figure 1 depicts a BMR that is south of the solar equator, has negative leading flux, and is greatly tilted away from the equator. To the left of the arrow, the schematic represents the BMR as it emerges and its two opposite-polarity flux domains are *slightly separated and growing in area and flux* (as in Figure 2 and its animation). To the right of the arrow, the schematic represents the BMR after it has finished emerging and *merging* of its opposite-polarity flux domains has made an in-sunspot sharp PIL between them. Only one of our 29 in-sunspot sharp PILs (number 12 in Table 1) has Type I genesis. Our schematic for Type I genesis is tailored to the genesis of that in-sunspot sharp PIL.

For Type II genesis, a mainly east-west BMR emerges east of another mainly east-west emerging BMR, and the in-sunspot sharp PIL is made by merger of the leading end of the eastern emerging BMR with the trailing end of the western emerging BMR. In the schematic for Type II genesis in Figure 1, to the left of the arrow, the two east-west pairs of opposite-polarity flux with negative leading flux represent the two mainly east-west emerging BMRs early in their emergence. To the right of the arrow, the schematic represents the two BMRs later in their emergence, when merger of the eastern BMR's leading flux and the western BMR's trailing flux has made an in-sunspot sharp PIL between them. Our schematic for Type II genesis is tailored to the genesis of in-sunspot sharp PIL number 2 in Table 1. Eleven of our 29 in-sunspot sharp PILs have Type II genesis.

In Type III genesis, the in-sunspot sharp PIL is made by the emergence of a relatively small BMR at an edge of a larger extant sunspot. The BMR's flux of the polarity of the sunspot can be partly or entirely inside the edge of the sunspot. In the schematic for Type III genesis in Figure 1, to the left of the arrow the schematic represents a relatively small BMR emerging at the north edge of a larger negative sunspot. The south edge of the BMR's positive flux is against the sunspot. The BMR's negative flux is in two roughly equal parts. One part is outside the edge of the sunspot, and the other part is right inside the edge. To the right of the arrow, the schematic represents the BMR later in its emergence, when much of its negative flux outside the sunspot has migrated farther westward several diameters of the BMR's positive flux. By then, merger of the BMR's positive flux both with the sunspot and with the BMR's negative flux inside the sunspot has made an in-sunspot sharp PIL between the BMR's positive flux and both the sunspot's negative flux and the BMR's negative flux that is against the south side of the BMR's positive flux. The schematic for Type III genesis is tailored to the genesis of the in-sunspot sharp PIL number 22 in Table 1. Five of our 29 in-sunspot sharp PILs have Type III genesis.

We define Type IV genesis of an in-sunspot sharp PIL to be any genesis that is not Type I, II, or III. An example of Type IV genesis is the genesis of sharp PIL number 17 in Table 1. That genesis is the merger of an emerging BMR with itself and with another emerging BMR. Another example is the genesis of sharp PIL 7. That genesis is the merger of two emerging BMRs with themselves and with larger unipolar flux. In the schematic for Type IV genesis in Figure 1, to the left of the arrow the schematic represents two emerging BMRs – both north of the solar equator, one east of the other – that have positive leading flux and are tilted away from the equator. The eastern emerging BMR is near the trailing (negative) end of the western emerging BMR. To the right of the arrow, the schematic represents the configuration later in the emergence of the two BMRs. By then, the eastern BMR has merged both with itself and with the trailing (negative) flux of the western BMR. That has made an in-sunspot sharp PIL between the merged positive and negative fluxes. The Figure 1 schematic for Type IV genesis is tailored to the genesis of in-sunspot sharp PIL 17. Twelve of our 29 in-sunspot sharp PILs have Type IV genesis.

In Appendix A we compare our four-category classification of delta-sunspot genesis to the three-category classification of H. Zirin & M. A. Liggett (1987). Their classification is for the genesis of delta sunspots that make major flares. Our classification is for the genesis of all sharp-PIL delta sunspots regardless of their productivity of major flares.

### 3.2. Example of Each Type of Genesis

#### 3.2.1. Example of Type I Genesis

Our example of Type I genesis is the observed genesis of in-sunspot sharp PIL number 12 in Table 1. In our search through HMI magnetograms and continuum images, that in-sunspot sharp PIL was found in the magnetogram and continuum image taken at 00:00 UT on 2021 September 26. It is the PIL of a small delta sunspot in AR 12876. AR 12876 is a small active region that emerges against the north side of an older active region, AR 12871, in the southern hemisphere. AR 12871 rotates onto the disk on September 18, and has negative leading polarity as is normal for active regions in the southern hemisphere in 2021

September, which is in the second year in the rise of Sunspot Cycle 25 (L. A. Upton & D. H. Hathaway 2023). AR 12876 is a BMR that begins emerging in the first hour or so of September 24. The BMR emerges for about 30 hours, during which the BMR's positive and negative flux domains are slightly separated, that is, the PIL between them is often not sharp (see Figure 2 and its animation). After emergence ends, some of the BMR's positive and negative flux gradually merges to make the delta sunspot that has the persistent sharp PIL seen in AR 12876 at 00:00 UT on September 26. The evolution of AR 12876 leading to the delta sunspot and its sharp PIL is shown in Figure 2 and its animation. The Figure 2 animation shows that the PIL becomes sharp by 17:00 UT on September 25 and is still sharp at the end of the animation, about 12 hours later.

In each of the three magnetograms in Figure 2, the yellow oval closely encloses nearly all of the BMR's positive and negative flux. In each oval, the area of the BMR's negative flux leads – is centered west of – the center of the area of the BMR's positive flux. That is, this active-region BMR, like the great majority of bipolar active regions in the southern hemisphere in Cycle 25, has negative leading flux. Hence, the BMR of AR 12876 is presumably the signature of a roughly east-west Ω-loop flux rope that has bubbled up and emerged from a band of westward pointing east-west magnetic field generated at the bottom of the solar convection zone by the sun's global dynamo process (e.g., C. Zwaan 1987; R. L. Moore, J. W. Cirtain & A. C. Sterling 2016; R. L. Moore, S. K. Tiwari, N. K. Panesar, & A. C. Sterling 2020; P. Charbonneau 2020; Y. Fan 2021).

By showing that, relative to each other, (A) the emerging BMR's negative flux and sunspot pores and (B) its positive flux and pores, both migrate counterclockwise, the Figure 2 animation shows that the emerging BMR pivots counterclockwise. We judge from the evolving tilt of the BMR from east-west that the BMR pivots counterclockwise more than 45° during the 42 hour span of the upper panel of Figure 2. That much counterclockwise pivot is indicated by the tilts of the long axes of the three yellow ovals in Figure 2. The tilt of each oval is roughly the BMR's tilt at that time. In the first magnetogram, the oval tilts more than 45° clockwise from east-west. In the second and third magnetograms, the ovals tilt 10° - 20° counterclockwise from east-west. On that basis, we are confident that the BMR of AR 12876 pivots more than 45° counterclockwise as it emerges.

Obvious counterclockwise pivot of an emerging BMR is a sign that the magnetic field of the emerging Ω-loop flux rope has definite right-handed twist about the flux rope's center field line (R. L. Moore, S. K. Tiwari, V. Aparna et al 2025). That the emerging BMR of AR 12876 pivots > 45° indicates (1) that the twist pitch angle of the field in the Ω loop's emerging top edge is > 45°, and (2) that the coronal field of the emerged bipolar active region (AR 12876) should display noticeable overall right-handed twist, i.e., noticeable overall S shape, in coronal images (R. L. Moore, S. K. Tiwari, V. Aparna, et al 2025). We judge from Figure 2 and its animation that, near and after the end of emergence of AR 12876, coronal EUV images from AIA's 211 Å channel show that the AR's coronal field does have overall S shape.

### 3.2.2. Example of Type II Genesis

Our example of Type II genesis is the observed genesis of in-sunspot sharp PIL number 2 in Table 1. In our search through full-disk HMI magnetograms and continuum images, that in-sunspot sharp PIL was found in the magnetogram and continuum image taken at 00:00 UT on 2014 September 28. It is the PIL of a delta sunspot in the middle of AR 12175, in the northern hemisphere. September 2014 is during the first year after the maximum of Sunspot Cycle 24 (L. A. Upton & D. H. Hathaway 2023). In Cycle 24, the great majority of active regions in the northern hemisphere have negative leading flux. AR 12175 is an overall bipolar active region having negative leading flux.

AR 12175 begins emerging east of central meridian during September 24, and rotates past central meridian during September 25. At 00:00 UT on September 26, two new BMRs having negative leading flux are emerging inside the overall BMR of AR 12175, one bigger than the other. The bigger new BMR is emerging in the trailing (eastern) half of AR 12175, and the smaller new BMR is emerging in the leading (western) half of AR 12175. At this time, the leading (negative) flux of the bigger new BMR and the trailing (positive) flux of the smaller new BMR are approaching each other and are near each other, but are not yet against each other. Six hours later those two fluxes are colliding as they continue emerging. By 18:00 UT on September 26, merger of the bigger new BMR's negative flux with the smaller BMR's positive flux has made a delta sunspot that has over twice more negative flux than positive flux. The delta sunspot's positive flux is against the south side of the delta sunspot's negative flux, and the PIL between them is sharp and engulfs the north half of the delta sunspot's positive flux. The delta sunspot continues to have that flux configuration 30 hours later, at 00:00 UT on September 28, the time at which this sharp PIL in AR 12175 was selected.

The evolution of AR 12175 leading to the delta sunspot and its sharp PIL is shown in Figure 3 and its animation. In the third magnetogram of Figure 3, the red arrow points to the delta sunspot's sharp PIL at the time of the sharp PIL's selection. In the first magnetogram of Figure 3, the two yellow ovals enclose the two new emerging BMRs before the negative flux of the eastern new BMR collides with the positive flux of the western new BMR. The second magnetogram of Figure 3 is after the negative flux of the eastern BMR and the positive flux of the western BMR have merged and have made the delta sunspot. In that magnetogram, the upward arrow points to the delta sunspot's positive flux, and the downward arrow points to the delta sunspot's negative flux.

The collision of the bigger new BMR's negative flux with the smaller new BMR's positive flux starts by 06:00 UT on September 26. Over the next 42 hours, from 06:00 UT on September 26 to 00:00 UT on September 28, those two opposite-polarity fluxes gradually press against each other and shear against each other. The positive flux continues its eastward migration and the negative flux continues its westward migration. Before 06:00 UT on September 26, as the negative leg of the bigger new BMR's emerged $\Omega$ loop and the positive leg of the smaller new BMR's emerged $\Omega$ loop approached each other, they presumably reconnected above the PIL between them, thereby making a magnetic arcade across the PIL. We expect the shearing of the positive flux and negative flux relative to each other to give the magnetic arcade left-handed shear. From close inspection of the third continuum image of Figure 3 and of the animation's continuum images for hours before, we discern faint light and dark striations in the sharp PIL's penumbral channel, and that these striations have left-handed shear across the PIL. Their left-handed shear is evidence that the magnetic field across the PIL has left-handed shear.

### *3.2.3. Example of Type III Genesis*

Our example of Type III genesis is the observed genesis of in-sunspot sharp PIL number 22 in Table 1. In our search of full-disk HMI magnetograms and continuum images, that in-sunspot sharp PIL was found in the magnetogram and continuum image taken at 18:00 UT on 2023 January 21. It is the PIL of a delta sunspot that has very much more negative flux than positive flux. Correspondingly, the delta sunspot's positive umbra is very much smaller than its negative umbra. The delta sunspot's positive flux is a small part of AR 13190. AR 13190 is an east-west overall bipolar active region in the southern hemisphere. It has negative leading flux, nearly all of which is in a big sunspot. January 2023 is roughly three years into the rise of Sunspot cycle 25 (L. A. Upton & D. H. Hathaway 2023). In Cycle 25, the great majority of active regions in the southern hemisphere have negative leading flux. The small positive flux domain of the extremely asymmetric delta sunspot is from a small east-west BMR that has negative leading flux and emerges against the north edge of the big negative sunspot.

AR 13190 rotates onto the disk on January 13 as a mature overall bipolar active region having negative leading flux. By January 15, the BMR is seen to be directed roughly east-west, has nearly all of its negative flux in a big unipolar sunspot, and has its positive (trailing) flux in plage in which there are a few small sunspots. From January 15 to January 20, the AR's big negative sunspot and its trailing positive flux keep their size and shape.

The small east-west BMR's emergence that makes the delta sunspot begins on January 20 at about 18:00 UT. By 00:00 UT on January 21, some of the emerging BMR's negative (leading) flux has merged with the northern fringe of the big negative sunspot, and the rest of the emerging BMR's negative flux is close to but outside the edge of the big sunspot and is migrating westward. By that time, the BMR's emergence against the big sunspot has made the extremely asymmetric delta sunspot and its sharp PIL. The PIL runs east-west through penumbra that links a small sunspot in the BMR' easternmost clump of positive flux to the north edge of the big sunspot's penumbra. From then (00:00 UT on January 21) through the time at which the delta sunspot's sharp PIL was selected (18:00 UT on January 21), the BMR continues to emerge. Some of the emerged negative flux is in the big sunspot's outer northern penumbra. The emerged negative flux outside the big sunspot rapidly migrates westward. The BMR's growing positive flux continues to be the positive flux of the extremely asymmetric delta sunspot and slowly migrates eastward.

The magnetic evolution in AR 13190 that makes the extremely asymmetric delta sunspot and the sharp PIL through it is shown in Figure 4 and its animation. In the third magnetogram of Figure 4, the red arrow points to the delta sunspot's sharp PIL at the time of the sharp PIL's selection. In the first magnetogram of Figure 4, the yellow oval encloses the emerging BMR early in its emergence, before some of its growing negative flux has merged with the northern fringe of the big negative sunspot's penumbra.

In the second magnetogram of Figure 4, the yellow oval encloses the BMR ten hours later. The animation shows that by then much of the BMR's emerging negative flux has merged with the northern fringe of the big sunspot's penumbra. The merger of the BMR with the big sunspot has resulted in both (1) continuous penumbra between the big negative sunspot and the BMR's growing small positive sunspot, and (2) the delta sunspot's sharp PIL. From the animation, we judge that the sharp PIL is between the BMR's positive flux and the part of the BMR's negative flux that has merged with the northern fringe of the big negative sunspot's penumbra.

At the time of the second magnetogram of Figure 4, the part of the BMR's negative flux that did not merge with the big sunspot is migrating westward much faster than the BMR's positive flux is migrating eastward. In the third magnetogram of Figure 4, the yellow oval encloses the BMR twelve hours later. By then, the leading end of the BMR's negative flux is more than twice farther from the BMR's positive flux than twelve hours earlier, and the BMR's positive flux has migrated farther east by roughly half it's east-west diameter. The eastward migration of the BMR's positive flux relative to the BMR's negative flux that is merged with the big sunspot's penumbra presumably imparts right-handed shear to the BMR's magnetic field crossing the sharp PIL. In the third continuum image of Figure 4, the red rectangle encloses penumbral fibrils that cross the sharp PIL at acute angles clockwise of the PIL. That is evidence that the BMR's magnetic field does have right-handed shear across the sharp PIL.

### 3.2.4. Example of Type IV Genesis

Our example of Type IV genesis is the observed genesis of in-sunspot sharp PIL number 17 in Table 1. In our search of full-disk HMI magnetograms and continuum images, that in-sunspot sharp PIL was found in the magnetogram and continuum image taken at 12:00 UT on 2022 May 19. It is the PIL of the big delta sunspot in AR 13014 in the northern hemisphere. 2022 May is in the third year of the rise of Sunspot Cycle 25 (L. A. Upton & D. H. Hathaway 2023). In Cycle 25, the great majority of active regions in the northern hemisphere have positive leading magnetic flux. AR 13014 has positive leading flux. At 12:00 UT on May 19, AR 13014 is centered about 10° east of central meridian, and the delta sunspot fills much of the trailing (eastern) half of AR 13014. The delta sunspot is the outcome of flux emergence that begins in AR 13014 three days earlier, by 00:00 UT on May 16.

AR 13014 rotates onto the disk on May 14. At 00:00 UT on May 15, AR 13014 is a slowly emerging BMR centered at (N21°, E70°). It has positive leading flux and is directed roughly east-west. Most of the AR's positive flux is in a leading sunspot.

By 06:00 UT on May 16, the BMR remains directed east-west, its leading sunspot is twice larger, and the BMR has completed its emergence. Also by then, a new BMR with positive leading flux is emerging in the south side of the middle of the emerged east-west BMR. The new BMR is tilted about 45° counterclockwise from east-west, has growing small sunspots in its leading (positive) and trailing (negative) flux domains, and its growing positive flux is approaching the southwest edge of the east-west BMR's positive sunspot. By 18:00 UT on May 16, the western edge of the emerging tilted BMR's leading (positive) flux touches the eastern edge of the east-west BMR's leading positive flux, and the tilted BMR has as much or more positive flux than the east-west BMR.

By 00:00 UT on May 17, the sunspot in the tilted BMR's positive flux has about twice the diameter of the east-west BMR's positive sunspot. By 06:00 UT on May 17, the tilted's positive sunspot is larger yet and is merging with the back (east) side of the east-west BMR's positive sunspot. Also by then, a second new tilted BMR with positive leading flux is emerging near the northeast side of the first tilted emerging BMR's trailing (negative) flux. The second new BMR is tilted about the same as the first new BMR, about 45° counterclockwise from east-west.

By 00:00 UT on May 18, the big leading positive sunspot of the first tilted BMR and the east-west BMR's leading sunspot are merged into one positive sunspot having two umbrae. The growing trailing (negative) flux of the first tilted BMR has sunspots and is colliding with and shearing against the southwest side of the second emerging tilted BMR. The second tilted BMR has a big growing sunspot in each polarity domain. The second tilted BMR's leading positive flux and first tilted BMR's trailing negative flux are shearing against each other. The sense of the shear is left-handed.

By 00:00 UT on May 19, the second tilted BMR's big leading positive sunspot and its big trailing negative sunspot have both merged with the first tilted BMR's big trailing negative sunspot and with each other. These three merged sunspots, together, are a big delta sunspot in which the PIL is sharp and is tilted roughly 20° counterclockwise from east-west. By 12:00 UT on May 19, the three sunspots composing the delta sunspot are packed together more tightly and the sharp PIL between them has a short new segment that is between the impacted opposite-polarity sunspots of the second new BMR.

After 18:00 UT on May 17, the magnetic evolution of AR 13014 that leads to the big delta sunspot and its sharp PIL at 12:00 UT on May 19 is shown in Figure 5 and its animation. In the third magnetogram of Figure 5, the western red arrow points to the sharp PIL's older and longer western segment, and the eastern red arrow points to the sharp PIL's newer and shorter eastern segment. In each of the three magnetograms, the first tilted BMR is inside the western yellow outline, and the second tilted BMR is inside the eastern yellow outline. Over the time interval (18:00 UT on May 17 to 12:00 UT on May 19) spanned by both Figure 5 and its animation, Figure 5 and its animation show that both tilted BMRs continue emerging while the second tilted BMR's growing big positive sunspot and growing big negative sunspot become packed together with each other and with the first tilted BMR's growing big negative sunspot. Their packing together makes the big delta sunspot and its sharp PIL.

The left-handed shearing of the first tilted BMR's negative flux against the southwest side of the second tilted BMR should be expected to give the delta sunspot's magnetic field left-handed shear across its PIL. In the third AIA 171 Å image in Figure 5, the two red arrows point to arch filaments that cross the sharp PIL at acute counterclockwise angles to the PIL. That is evidence that the delta sunspot's magnetic arcade low across the sharp PIL does have left-handed shear.

### *3.3. Table 1*

The key observed aspects of each of our 29 selected sharp PILs in sunspots and its genesis are listed in Table 1. The first column of Table 1 gives the chronological ordinal number of each in-sunspot sharp PIL. The NOAA AR number of the active region in which the in-sunspot sharp PIL is made is in the second column. The third, fourth, and fifth columns give the date, time, and (x, y) coordinates of the in-sunspot sharp PIL when it was selected. The sixth column states whether the selected in-sunspot sharp PIL is the PIL of a delta sunspot. The seventh column states whether the selected in-sunspot sharp PIL is made by the evolution of only a single bipolar magnetic region (BMR). The last column (eighth column) gives the in-sunspot sharp PIL's genesis Type.

The main take-away from Table 1 is that only one of the 29 in-sunspot sharp PILs (number 12) is made from a *single* BMR. Each of the other 28 sharp PILs in sunspots results from the merger of two or more BMRs.

## 4. Interpretation

### *4.1. Type I Genesis*

Type I genesis of an in-sunspot sharp PIL is genesis by inside merging of a single BMR's opposite-polarity flux (Figure 1). For the cause of the inside merging we have two alternative scenarios for the evolution of the BMR's flux-rope loop: scenario I-A and scenario I-B. Figure 6a is a schematic depiction of scenario I-A, and Figure 6b is a schematic depiction of scenario I-B. Each of these is tailored to the magnetic evolution of AR 12876 shown in Figure 2 and its animation. That evolution makes in-sunspot sharp PIL number 12 in Table 1. Of our 29 random in-sunspot sharp PILs, sharp PIL 12 is the only one that has Type I genesis. (In our schematic drawings in Figures 6a, 6b, 7, 8, and 9, realistic lateral and vertical expansion of the emerged magnetic field above the photosphere is ignored. Only the field's essential topology is depicted.)

Scenario I-A is depicted in Figure 6a by a sequence of four schematic drawings of an upper extent of a flux-rope Ω loop before, during and after its partial emergence through the photosphere. The depicted Ω-loop flux rope is directed solar east-west and has negative leading flux. Following R. L. Moore, S. K. Tiwari, N. K. Panesar, & A. C. Sterling (2020), we assume the Ω loop has bubbled up from a band of west-pointing horizontal magnetic field generated at the bottom of the convection zone by the Sun's global dynamo process. (AR 12876 has negative leading flux and emerges in the southern hemisphere in Sunspot Cycle 25. In Cycle 25 nearly all southern-hemisphere active regions have negative leading flux. That implies that the magnetic field in the dynamo-generated horizontal band of field in the southern hemisphere points west when AR 12876 emerges, e.g., P. Charbonneau 2020).

In the drawings in Figure 6a, the Ω loop is centered on a down-flow between two convection cells that each have horizontal spans as large or larger than that of the Ω loop. We suppose the Ω loop grew from a slight convex-upward bend in a horizontal flux tube deep in the convection zone a la R. L. Moore, S. K. Tiwari, N. K. Panesar, & A. C. Sterling (2000). As in their scenario for making flux-rope Ω loops, we suppose the flux tube's initial upward bend is made by up-flow in the bottom of a convection cell. For our scenario depicted in Figure 6a, that convection cell is similarly large as the two convection cells depicted in Figure 6a, sits in the foreground south of them, and abuts them and the down-flow between them. The converging outflows of the two depicted cells and the northward outflow of the southern cell flow into the depicted central down-flow. We suppose that by the time the of rising Ω loop is near the photosphere, as in the top-left drawing, the southern convection cell's northward outflow into the down-flow has swept the top of the Ω loop into the down-flow. We further suppose that the top of the Ω loop is buoyant enough to overcome the down-flow's downward pull enough to partly emerge through the photosphere and make a BMR such as the BMR of AR 12876 in Figure 2 and its animation.

In the drawings in Figure 6a, the magnetic field of the flux rope has right-handed twist about the flux rope's centerline. That is appropriate for an emerging BMR Ω loop in the southern hemisphere that makes an S-shape sigmoidal active region such as AR 12876 (R. L. Moore, S. K. Tiwari, V. Aparna, et al. 2025). We assume that the pitch of the twist in the flux rope's magnetic field decreases monotonically inward to zero at the flux rope's centerline. As a result, as the top-left, top-right, and bottom-left drawings together depict, (1) when the top edge of the Ω loop first emerges the emerged bipole (viewed from above) has negative flux leading and is tilted clockwise from east-west, and (2) as the rest of the upper half of the Ω loop's flux emerges, the emerging bipole decreases its tilt, i.e., pivots counterclockwise to become directed more nearly east-west (R. L. Moore, S. K. Tiwari, V. Aparna, et al. 2025).

The top-right and bottom-left drawings in Figure 6a depict that as the top part of the Ω loop emerges, the emerging BMR's opposite-polarity flux initially spreads slightly apart, as is the case for the emerging BMR of AR 12876 in Figure 2 and its animation. The bottom-right drawing in Figure 6a depicts that we assume the inflow to the down-flow eventually overcomes the Ω loop's buoyancy, stops the Ω loop's emergence, and then draws together some of the emerged flux to make, on the down-flow, a sharp-PIL delta sunspot like that in the third magnetogam and continuum image in Figure 2.

Our alternative scenario (scenario I-B) for the cause of the single BMR's inside merging that makes the sharp-PIL delta sunspot in AR 12876 is depicted in Figure 6b. In that scenario, we assume that the BMR results from the partial emergence of the top of a writhe-made kink in an initially horizontal east-west flux rope in the convection zone. This aspect of the scenario is the same as in the scenario for the genesis of delta sunspots proposed by K. Tanaka from observations, and by Zweibel, M. G. Linton, & G. Fisher (1999) from MHD modeling. The twist of the field in the initial (pre-kinked) flux rope is assumed to have been great enough to have made some of the flux rope's twist go into the writhe kink.

Scenario I-B has two basic differences from scenario I-A. One is that whereas the magnetic field in the initial pre-Ω-loop horizontal flux tube in scenario I-A points west, the magnetic field in the initial pre-kinked horizontal flux rope in scenario I-B points east. The other basic difference is that whereas the BMR in scenario I-A is made by the emergence of the top of a flux-rope Ω loop, the BMR in scenario I-B is made by the emergence of the top of a flux rope's writhe kink.

Except for the above two differences, scenario I-B is the same as scenario I-A. In each, prior to emerging, the top of the flux-rope loop is swept into down-flow at the north edge of a large convection cell south of the downflow by that cell's northward flow into the down-flow. In each, the buoyancy of the top of the flux-rope loop at first overcomes the downward pull of the down-flow and emerges to make a BMR whose two opposite-polarity flux domains grow and initially spread slightly apart. In each, the emerging BMR pivots counterclockwise and gives the active region's field in the corona overall S shape. In each, the down-flow and the inflow to the down-flow eventually halt the emergence of the flux-rope loop top, and pack some of the BMR's opposite-polarity flux together to make a sharp-PIL delta sunspot that sits on the down-flow.

In scenario I-B the eastward direction of the field in the initial flux rope is opposite the westward direction that the field in the initial flux ropes evidently has for nearly all the active regions in the southern hemisphere when AR 12876 emerges. For that reason, we think scenario I-A is more plausible than scenario I-B. That is, we think the BMR for AR 12876 is more plausibly the top of an east-west Ω-loop flux rope having right-handed twist than an east-west top of a kink in a flux rope having right-handed twist.

## 4.2. Type II Genesis

Type II genesis of an in-sunspot sharp PIL is genesis by outside merging of the leading flux of a mainly east-west emerging BMR with the opposite-polarity trailing flux of another mainly east-west emerging BMR (Figure 1). Our scenario for the cause of the outside merging is schematically depicted in Figure 7. The depiction is tailored to the magnetic evolution of AR 12175 shown in Figure 3 and its animation. That evolution makes in-sunspot sharp PIL number 2 in Table 1. Of our 29 random in-sunspot sharp PILs, 11 have Type II genesis like that of in-sunspot sharp PIL 2.

Figure 7 is a sequence of three schematic drawings of upper extents of two head-to-tail Ω loops before, during, and after their emergence through the photosphere. Each Ω loop is directed east-west and has negative leading flux. As we assume for the origin of the rising flux-rope loop in our scenario I A for Type I genesis, we assume that each of these two Ω loops has bubbled up from a band of west-pointing horizontal magnetic field generated at the bottom of the convection zone by the Sun's global dynamo process. (AR 12175 has negative leading flux and emerges in the northern hemisphere in Sunspot Cycle 24. In Cycle 24 nearly all northern-hemisphere active-region BMRs have negative leading flux. That implies that the magnetic field in the dynamo-generated horizontal band of field in the northern hemisphere points west when AR 12175 emerges.)

In Figure 7, the top of each Ω loop is centered on the up-flow of one of two adjacent convection cells that are east-west of each other. The westward outflow of the eastern cell and the eastward outflow of the western cell feed the middle down-flow between them. The eastern cell's eastward outflow feeds the down-flow on that cell's eastern edge, and the western cell's westward outflow feeds the down-flow on that cell's western edge.

In the top panel of Figure 7, the top edge of each Ω loop has risen to just below the photosphere. We suppose that by that time, far below the convection zone's upper layer depicted in Figure 7, the outflows of the two convection cells have swept each of the four legs of the two Ω loops into the nearest down-flow. In the middle panel of Figure 7, most of the flux in each Ω loop has emerged through the photosphere. The Ω loop's flux in photospheric magnetograms is a BMR that is nearly finished emerging and has its two opposite-polarity flux domains spreading apart. The negative flux of the eastern BMR and the positive flux of the western BMR are converging but are not yet touching, as in the first magnetogram and continuum image of AR 12175 in Figure 3 and early in the animation. Also in the middle panel of Figure 7, the emerged positive flux of the eastern BMR and the emerged negative flux of the western BMR are each being swept into the nearest downflow.

The bottom panel of Figure 7 depicts the endpoint of our scenario for Type II genesis of in-sunspot sharp PILs. By now, the two converging legs of the two emerged Ω loops have been swept into the middle down-flow. The eastern foot of the eastern emerged Ω loop has become centered on the eastern down-flow, and the western foot of the western emerged Ω loop has become centered on the western downflow. We suppose the packing together of the western BMR's positive flux and eastern BMR's negative flux by the inflow to the middle down-flow makes on the down-flow a sharp-PIL delta sunspot like the delta sunspot in AR 12175 in the second and third magnetograms and continuum images in Figure 3 and in the animation.

### 4.3. Type III Genesis

Type III genesis of an in-sunspot sharp PIL is genesis by emergence of a BMR at an edge of a unipolar larger sunspot (Figure 1). Our scenario for the cause of Type III genesis is schematically depicted in Figure 8. With a major alteration, the schematic drawings in Figure 8 are tailored to the magnetic evolution in and around the north edge of the big negative sunspot that is the leading sunspot of AR 13190, the magnetic evolution shown in Figure 4 and its animation. That evolution makes in-sunspot sharp PIL number 22 in Table 1. Of our 29 random in-sunspot sharp PILs, five have Type III genesis essentially like that of in-sunspot sharp PIL 22.

In the case of the genesis of in-sunspot sharp PIL 22, much of the emerging BMR's flux of the polarity of the big sunspot (negative in this case) emerges outside the big sunspot and migrates westward. The outside of the rest of the BMR's negative flux melds with the north edge of the big sunspot's penumbra, and the inside of that remainder of the BMR's negative flux melds with the BMR's positive flux to make the sharp PIL. The overall emerging BMR is evidently tilted clockwise of east-west.

In other cases of Type III genesis, if the emerging BMR's flux of the big sunspot's polarity is taken to be the BMR's head, then, compared to the Type III genesis of sharp PIL 22, the BMR can emerge either more head-on to the big sunspot or more tail-on. To enable the drawings to be in a vertical plane, even though the schematic drawings in Figure 8 are tailored to the magnetic evolution shown in Figure 4 and its animation, the depicted evolution is for the corresponding case of the BMR emerging tail-on to the big sunspot at the west edge of the big sunspot, instead of for the partly head-on emergence of the emerging BMR at the north edge of the big sunspot in Figure 4. More generally, the drawings in Figure 8 depict the corresponding case in which the BMR's emerging flux of polarity opposite that of the big sunspot emerges straight against the big sunspot and the BMR's emerging flux of the big sunspot's polarity migrates straight away from the big sunspot.

Figure 8 is a sequence of three schematic drawings depicting an edge of a big negative sunspot and an upper extent of a flux-rope Ω loop before, during, and after the top of the Ω loop emerges with its positive flux straight against the sunspot and its negative flux straight away from the sunspot. As in Figures 6 and 7, the curves are in a vertical plane, the red curves are magnetic field lines, and the blue curves are streamlines of subphotospheric convection. Here, the vertical plane bisects the Ω loop and the sunspot.

In the top drawing in Figure 8, the rising Ω loop is centered in the up-flow of a convection cell that has its down-flow on one side against the outer edge of the sunspot's magnetic field sheaf below the photosphere and has its down-flow on its opposite side near the Ω loop's leg away from the sunspot. The top edge of the Ω loop is about to emerge through the photosphere, and each Ω-loop leg is being swept into the convection cell's nearest down-flow.

In the middle drawing in Figure 8, about half of the Ω loop's flux has emerged. The outside of the emerged field is reconnecting with the sunspot magnetic field's overlying canopy. That reconnection builds a magnetic arcade over the PIL between the BMR's positive flux and the sunspot's negative flux. That PIL is viewed end-on in the middle and bottom drawings in Figure 8, and is marked by a vertical red tick mark on the top of the photosphere. The PIL in these drawings is the analogue of the in-sunspot sharp PIL in the greatly asymmetric delta sunspot seen in the second and third magnetograms and continuum images in Figure 4 and in Figure 4 animation. In the middle drawing in Figure 8, each leg of the emerging $\Omega$ loop continues to be swept into the nearest down-flow. In addition, the convection cell's downflow farther from the sunspot is migrating away from the sunspot and is pulling the $\Omega$ loop's leg with it. That depicts our scenario for why the BMR's negative flux migrates westward between the first and second magnetograms in Figure 4.

The bottom drawing in Figure 8 depicts the endpoint of our scenario for Type III genesis of in-sunspot sharp PILs. In that drawing, the outer field of the emerged $\Omega$ loop continues to reconnect with the sunspot's canopy field, adding more loops to the magnetic arcade enveloping the PIL between the BMR's positive flux and the sunspot's negative flux. By this time, we suppose the convection cell's eastward outflow into the eastern down-flow has packed the BMR's positive flux strongly enough against the sunspot's negative flux to make an in-sunspot sharp PIL like that in the third magnetogram in Figure 4. Also in the bottom drawing, the convection cell's down-flow farther from the sunspot is farther from the sunspot than in the middle drawing and has pulled the BMR's negative flux farther into it. That depicts our scenario for the analogous further westward migration of the BMR's negative flux between the second and third magnetograms in Figure 4.

### 4.4. Type IV Genesis

One variety of Type IV genesis of in-sunspot sharp PILs is genesis by the opposite-polarity fluxes of an emerging BMR merging with each other and with another BMR's flux of one polarity (Figure1). Our scenario for the cause of that variety of Type IV genesis is schematically depicted in Figure 9. With a major alteration, the schematic drawings in Figure 9 are tailored to the magnetic evolution of AR 1304 shown in Figure 5 and its animation. That evolution makes in-sunspot sharp PIL 17 in Table 1.

In-sunspot sharp PIL 17 is the PIL of the big delta sunspot in AR 13014 on 2022 May 19. Figure 5 and its animation show the delta sunspot and its sharp PIL result from the opposite-polarity fluxes of an emerging BMR merging against each other and against the negative flux of another emerging BMR. These two BMRs and AR 13014 in which they emerge have positive leading flux. The two BMRs are directed mainly east-west, have about the same clockwise tilt from east-west, and emerge in the northern hemisphere during Solar Cycle 25. In Cycle 25 nearly all northern-hemisphere emerging BMRs as large as the two emerging BMRs in AR 13014 in Figure 5 have positive leading flux. That implies the magnetic field in the dynamo-generated band of field in the northern hemisphere points east when AR 13014 emerges and has its big delta sunspot. We therefore assume that each of the two emerging BMRs that merge to make the big delta sunspot is a flux-rope $\Omega$ loop that has bubbled up from dynamo-generated east-pointing field deep in the convection zone.

To enable the drawings to be in a vertical plane through both emerging $\Omega$ loops, even though the schematic drawings in Figure 9 are tailored to the magnetic evolution shown in Figure 5 and its animation, the depicted evolution is for the corresponding case of the eastern BMR and the western BMR both having no tilt from east-west and coming straight together, head-do-tail, instead of the eastern BMR shearing against the north side of the western BMR's trailing (negative) flux as in Figure 5 and its animation.

Figure 9 is a sequence of three schematic drawings depicting upper extents of two east-west $\Omega$ loops before, during, and after they emerge. In the top drawing, each $\Omega$ loop is centered on the up-flow of a convection cell of its size, the east leg of each $\Omega$ loop has been convectively swept into the eastern down-flow of its $\Omega$-loop's convection cell, and the west leg of each $\Omega$ loop has been swept into the western down-flow of its $\Omega$ loop's convection cell. Each $\Omega$ loop is rising, and its top edge is about to start emerging through the photosphere, up through the chromosphere, and into the corona.

During the time between the top and middle drawings in Figure 9, we suppose the convection cell on which the eastern $\Omega$ loop is centered in the top drawing is replaced by a western part of a larger convection cell that, in the middle and bottom drawings, flows into its western down-flow between it and the western $\Omega$ loop's convection cell. The middle and bottom drawings depict that we suppose the western $\Omega$ loop's convection cell does not change in the time between the middle and bottom drawings.

In the second drawing in Figure 9, nearly all of the flux in each Ω loop has emerged. The eastern Ω loop's negative flux is being swept toward that Ω loop's positive flux. The eastern Ω loop's positive flux and the western Ω loop's negative flux are being swept toward each other to meet over the eastern downflow. The western Ω loop's positive flux is being swept toward the western downflow.

The bottom drawing in Figure 9 depicts the endpoint of our scenario corresponding to the Type IV genesis of the two sharp-PIL segments pointed to by the two red arrows in the third magnetogram in Figure 5. In that drawing, the eastern Ω loop's positive flux and the western Ω loop's negative flux have been crammed against each other over the eastern down-flow. That made between them a sharp pill that is viewed end-on in the bottom drawing. That sharp PIL corresponds to the sharp-PIL segment pointed to by the western red arrow in the third magnetogram in Figure 5. Reconnection at the interface of the two emerged Ω loops has built a magnetic arcade enveloping the sharp PIL between the two Ω loops.

In the bottom drawing in Figure 9, the eastern Ω loop's negative flux is crammed against the eastern Ω loop's positive flux. That makes another sharp PIL that is viewed end-on in the bottom drawing. That sharp PIL corresponds to the sharp-PIL segment pointed to by the eastern red arrow in the third magnetogram in Figure 5. Reconnection at the inside interface of the eastern Ω loop's two crammed-together legs has built a magnetic arcade that straddles the eastern sharp PIL.

## 5. Summary and Conclusion

To explore how the Sun makes delta sunspots, we searched through years of full-disk magnetograms and simultaneous full-disk continuum images (from SDO/HMI) for sunspots that have in them a sharp PIL that is made while the sunspot's active region is well on the face of the Sun. By using Helioviewer to step through 10.7 years of those observations (2013 November through 2024 June), we found 29 such sharp-PIL sunspots. All of them but one are delta sunspots, i.e., have umbra on both sides of the PIL. The exception has umbra and penumbra on one side of its PIL, but only penumbra on the other side.

The 29 in-sunspot sharp PILs have four classes of genesis which we have named Type I, Type II, Type III, and Type IV. Type I is genesis by inside merging of a single BMR's opposite-polarity flux. Type II is genesis by outside merging of the leading end of an emerging east-west BMR with the opposite-polarity trailing end of another emerging east-west BMR. Type III is genesis by a BMR emerging at an edge of a unipolar larger sunspot. Type IV is any genesis that is not Type I, II, or III. Of the 29 in-sunspot sharp Pils, only one has Type I genesis, 11 have Type II genesis, 5 (including the sharp PIL in the non-delta sunspot) have Type III genesis, and 12 have Type IV genesis.

For each type of genesis, we present in detail an observed example. For each of these four observed-genesis examples we schematically depict our scenario or scenarios for the cause of that example genesis. The core idea in each scenario is that the sunspot and its sharp PIL sit on a convection down-flow because they are the result of opposite-polarity legs of one or two emerging (or emerged) flux-rope Ω loops being pushed together by the convection inflow to that down-flow.

For our only observed genesis of an in-sunspot sharp PIL from a single BMR Ω loop (Type I genesis), we present two alternative scenarios for the cause of the delta sunspot and its sharp PIL. In both scenarios, the sharp-PIL delta sunspot sits on down-flow between large convection cells and is made by the inflow to the down-flow packing together flux from the opposite-polarity feet of an emerged flux-rope Ω loop. In the first scenario, the flux-rope loop is the emerged top of a flux-rope Ω loop in which the horizontal direction of the magnetic field is that of nearly all of the BMRs in the sunspot cycle and hemisphere of the observed delta sunspot. In the second scenario, the flux-rope loop is the emerged top of a writhe kink of a sufficiently twisted flux rope. Because the horizontal direction of the magnetic field of the prospective writhe-kinked flux rope is opposite that of nearly all BMRs in the sunspot cycle and hemisphere of the observed delta sunspot, we judge that for this delta sunspot the first scenario is more likely than the second scenario.

Of our 28 sharp-PIL delta sunspots 27 are evidently not made by the emergence of a writhe kink in a sufficiently twisted subsurface horizontal flux rope. In addition, we reason that our one candidate delta sunspot for being made that way is plausibly made from the emerged top of a not-writhe-kinked Ω loop.

In these respects, our set of 28 sharp-PIL delta sunspots indicates that few, if any, sharp-PIL delta sunspots are the signature of an emerged writhe kink in a sufficiently twisted subsurface horizontal flux rope as proposed by K. Tanaka (1991) and Y. Fan, E. G. Zweibel, M. G. Linton, & G. H. Fisher (1999). Instead, our 28 delta sunspots (1) indicate that, except for at most a small fraction, sharp-PIL delta sunspots are made by convection flows packing together opposite-polarity feet of two or more emerging (or emerged) $\Omega$ loops, and (2) suggest that many of the rest might not be the signature of the emerged top of a flux-rope writhe kink either, but instead are made by convection flows packing together flux from the opposite-polarity feet of a not-kinked ordinary single emerged $\Omega$ loop.

The main conclusion of our paper is not that flux-rope writhe and kink have a negligible role in the making of delta sunspots. The paper does not rule out that one or more of the bipoles that merge to make a Type IV delta sunspot can be kinked or partially writhed flux ropes. However, the genesis of none of our 12 Type IV delta sunspots clearly suggests that possibility.

## Acknowledgements

This work was supported by the Heliophysics Division of NASA's Science Mission Directorate through the Heliophysics Guest Investigators (HGI) program and the Heliophysics Supporting Research (HSR) program. SKT, RLM, VA, and NKP sincerely acknowledge support from NASA HSR grant (80NSSC23K0093) and/or NSF AAG award (no. 2307505). SKT also acknowledges support from NSF SHINE (award no. 2230633), NASA HGI (80NSSC24K0551), and ARC-CREST (NASA Cooperative Agreement 80NSSC23M0230). VA Supported through the state of New Mexico and NSF SHINE. NKP acknowledges support from NASA's SDO/AIA (NNG04EA00C) grant and NASA's HSR (80NSSC24K0258) grant. We acknowledge use of SDO/AIA/HMI data. AIA is an instrument onboard the Solar Dynamics Observatory, a mission for NASA's Living With a Star program.

## APPENDIX A

## A COMPARISON OF OUR DELTA-SUNSPOT GENESIS TYPES WITH THOSE OF H. ZIRIN & M. A. LIGGETT (1987)

The well-known classical paper, "Delta Sunspots and Great Flares" by H. Zirin and M. A. Liggett (1987) classifies delta sunspots that have major flares into three distinct classes or types according to their genesis (Here, the letter p stands for the preceding (western) polarity of a BMR (M. J. Martres & A. Bruzek 1977).):

Type 1. "A complex of sunspots emerging all at once with different dipoles intertwined. We call those the "island" $\delta$ configurations. They are quite compact, with a large umbra, usually p polarity, surrounded by bright H$\alpha$ emission."

Type 2. "A single $\delta$ group, as Type 1, but formed in two stages, with large satellite dipoles emerging in the penumbra of a large p spot.

Type 3. "A $\delta$ configuration formed by collision between two separate but growing bipolar spot groups, the leader of one colliding with the follower of the other."

Citing H. Kunzel, W. Mattig, & E. H. Schroter (1961), H. Zirin & M. A. Liggett (1987) state that Type-3 $\delta$ configurations are "not so active" compared to Type-1 and Type-2 $\delta$ configurations. We take that to mean Type-1 and Type-2 $\delta$ configurations are more productive of major flares than are Type-3 $\delta$ configurations.

The above delta-sunspot-genesis classification by H. Zirin & M. A. Liggett (1987) is for δ configurations that produce major flares. In contrast, our classification of delta-sunspot genesis into four types (Type I, Type II, Type III, Type IV) is for all sharp-PIL delta sunspots, regardless of their productivity of major flares. We have not studied the flare productivity of any of our 28 delta sunspots. That aspect of our 28 delta sunspots is beyond the scope of this paper.

Comparing the definitions of delta-sunspot genesis Types I, II, II, IV for all delta sunspots regardless of their productivity of major flares (see Figure 1), with the above definitions of genesis Types 1, 2, 3 for delta sunspots that produce major flares, we conclude the following. Our genesis Type IV includes genesis Types 1 and 2, and our genesis Type II is the same as genesis Type 3. In addition, we surmise that due to the rarity of Type I genesis and perhaps due to low productivity of major flares by Type-I-genesis and Type-III-genesis delta sunspots, genesis Types I and III are not included in genesis Types 1, 2, 3. That is, the genesis of our 28 delta sunspots is entirely covered by the classification categories Type I, Type II, Type III, Type IV, but not entirely by the classification categories Type 1, Type 2, Type 3.

| | | | | | | | |
|---|---|---|---|---|---|---|---|
| **Table 1. Observed Aspects of the Genesis of 29 Sharp PILs in Sunspots** | | | | | | | |
| Ord. No. | NOAA AR No. | Selected PIL | | | Delta Sunspot? (Yes or No) | Single BMR? (Yes or No) | Genesis Type (I, II, II, IV) |
| | | Date (YYYY/MM/DD) | Time (UT) | Place (x", y") | | | |
| 1 | 11900 | 2013/11/15 | 06:00 | 255, -377 | Yes | No | **Type II** |
| 2 | 12175 | 2014/09/28 | 00:00 | 602, 162 | Yes | No | **Type II** |
| 3 | 12241 | 2014/12/19 | 06:00 | -114, -169 | Yes | No | **Type II** |
| 4 | 12242 | 2014/12/19 | 00:00 | 265, -296 | Yes | No | **Type II** |
| 5 | 12297 | 2015/03/13 | 12:00 | -188, -170 | Yes | No | **Type III** |
| 6 | 12434 | 2015/10/19 | 12:00 | -15, -249 | Yes | No | **Type III** |
| 7 | 12473 | 2015/12/26 | 18:00 | -196, -292 | Yes | No | **Type IV** |
| 8 | 12497 | 2016/02/12 | 12:00 | 253, 323 | Yes | No | **Type II** |
| 9 | 12465 | 2017/04/03 | 00:00 | 352, -79 | Yes | No | **Type IV** |
| 10 | 12673 | 2017/09/05 | 00:00 | 238, -261 | Yes | No | **Type IV** |
| 11 | 12860 | 2021/08/28 | 18:00 | 98, -521 | Yes | No | **Type IV** |
| 12 | 12876 | 2021/09/26 | 00:00 | 186, -538 | Yes | Yes | **Type I** |
| 13 | 12929 | 2022/01/17 | 12:00 | 589, 188 | No | No | **Type III** |
| 14 | 12975 | 2022/03/30 | 00:00 | 380, 317 | Yes | No | **Type IV** |
| 15 | 13004 | 2022/05/05 | 12:00 | 372, -196 | Yes | No | **Type IV** |
| 16 | 13007 | 2022/05/12 | 00:00 | -481, -336 | Yes | No | **Type IV** |
| 17 | 13014 | 2022/05/19 | 12:00 | -195, 383 | Yes | No | **Type IV** |
| 18 | 13051 | 2022/07/09 | 10:00 | 378, 396 | Yes | No | **Type III** |
| 19 | 13058 | 2022/07/17 | 12:00 | -511, 182 | Yes | No | **Type II** |
| 20 | 13078 | 2022/08/15 | 12:00 | 0, -473 | Yes | No | **Type IV** |
| 21 | 13089 | 2022/08/29 | 18:00 | -182, -447 | Yes | No | **Type II** |
| 22 | 13190 | 2023/01/21 | 18:00 | 447, -142 | Yes | No | **Type III** |
| 23 | 13236 | 2023/02/25 | 18:00 | 39, -299 | Yes | No | **Type II** |
| 24 | 13296 | 2023/05/09 | 06:00 | 420, 280 | Yes | No | **Type IV** |
| 25 | 13451 | 2023/10/08 | 00:00 | 494, 189 | Yes | No | **Type II** |
| 26 | 13599 | 2024/03/09 | 00:00 | 280. -120 | Yes | No | **Type IV** |
| 27 | 13663 | 2024/05/03 | 18:00 | 21, 472 | Yes | No | **Type II** |
| 28 | 13664 | 2024/05/05 | 00:00 | -403, -283 | Yes | No | **Type IV** |
| 29 | 13716 | 2024/06/20 | 00:00 | 472, 123 | Yes | No | **Type II** |

Figure captions

Figure 1. Definition and schematic of each of our four types of genesis of in-sunspot sharp PILs.

Figure 2. The only example of Type I genesis among the 29 in-sunspot sharp PILs in our sample. Upper panels are co-temporal HMI magnetograms, HMI continuum images, and AIA 211 Å images (all having the same field of view) at three stages of emergence and evolution of AR 12876 in 2021 September. Here, and in Figures 2 – 5, solar north is up and west is to the right, and the day and time are in each continuum image. Yellow ovals show the BMR's extent and tilt. The red arrow points to the delta sunspot's sharp PIL. In the second and third AIA panels, the dashed curve tracks the overall S shape of the AR's coronal field. The bottom panel is the time profile of the unsigned magnetic flux in the image panels' field of view. The accompanying 21s animation spans 54 hr (2021 September 24, 00:00 UT to September 26, 06:00 UT), has 15-min cadence, and has this figure's field of view.

Figure 3. Example of Type II genesis of an in-sunspot sharp PIL. Shown are co-temporal HMI magnetograms (top) and HMI continuum images (bottom) of AR 12175 at three stages of its evolution in 2014 September. In the first magnetogram, yellow ovals outline two emerging BMRs that subsequently merge head-to-tail. A day later, in the second magnetogram, the negative head (pointed to by the downward yellow arrow) of the eastern BMR has merged with the positive tail (pointed to by the upward yellow arrow) of the western BMR, making a delta sunspot. Another day later, in the third magnetogram the red arrow points to the delta sunspot's sharp PIL at the time of selection. In the third continuum image, penumbral striations cross that PIL at an acute counterclockwise angle, showing left-handed magnetic shear. The accompanying 19 s animation spans 48 hr (2014 September26, 00:00 UT to September 28, 00:00 UT), has 15-min cadence, and has this figure's field of view.

Figure 4. Example of Type III genesis of an in-sunspot sharp PIL. Shown are co-temporal HMI magnetograms (top) and HMI continuum images (bottom) at three stages during the emergence of a small BMR at the northern edge of the big negative leading sunspot of AR 13190 in 2023 January. In the first magnetogram, the yellow oval outlines the BMR early in its emergence. Ten hours later, in the second magnetogram, the BMR's positive flux has migrated eastward, part of the BMR's negative flux has merged with the big negative sunspot, and the rest of the BMR's negative flux has migrated westward. Now there is a sharp PIL between the BMR's positive flux and both some of the big sunspot's negative flux and the BMR's negative flux that has merged with the big sunspot. In the third continuum image, the red box spans penumbral striations that cross the PIL at an acute clockwise angle, showing right-handed magnetic shear. The accompanying 8 s animation spans 22 hr (2023 January 20, 20:00 UT to January 21, 18:00 UT), has 15-min cadence, and has this figure's field of view.

Figure 5. Example of Type IV genesis of an in-sunspot sharp PIL. Shown are co-temporal HMI magnetograms, HMI continuum images, and AIA 171 Å images at three stages of the evolution of AR 13014 in 2022 May. Yellow closed curves outline two BMRs that emerge and merge within the AR. In the first column, the western emerging BMR's leading positive sunspot has merged with the AR's preexisting leading positive sunspot, and the eastern BMR is emerging close behind (west of) the western emerging BMR's trailing negative flux. Eighteen hours later, in the second column, the eastern BMR has wrapped around the western BMR's trailing negative sunspot, making a large delta sunspot in which the PIL is not yet sharp. By the third column, at the PIl's selection time 24 hours later, the three merging sunspots are larger and the PIL has two sharp segments. The western segment is between the eastern BMR's positive sunspot and the western BMR's negative sunspot; the eastern segment is between the eastern BMR's positive sunspot and negative sunspot. In the magnetogram, red arrows point to the segments. In the AIA 171 Å image, red arrows point to striations crossing the two sharp-PIL segments at acute counterclockwise angles, showing left-handed magnetic shear. The accompanying 16 s animation spans

42 hr (2022 May 17, 18:00 UT to May 19, 12:00 UT), has 15-min cadence, and has this figure's field of view.

Figure 6a. Schematic of Scenario I-A for Type I genesis of the sharp-PIL delta sunspot in AR 12876 from a single east-west $\Omega$-loop flux rope without writhe. The thick black horizontal line is the photosphere. Blue curves are subphotospheric convection streamlines in a vertical east-west plane that is viewed from the south and bisects the flux rope. Red curves are projections of the flux rope and two key flux-rope field lines: the center field line lying in the plane and a surface field line having right-handed twist. Plus and minus signs label the $\Omega$ loop's positive-polarity foot and negative-polarity foot. During its emergence and partial submergence, the $\Omega$ loop is horizontally trapped in downflow between convection cells. Because the flux-rope field's twist is right-handed and its twist pitch decreases inward to zero at the center field line, the $\Omega$ loop's BMR footprint pivots counterclockwise in photospheric magnetograms and continuum images as the $\Omega$ loop emerges and the $\Omega$ loop's coronal field becomes an S-shaped sigmoid, as observed for AR 12876 (Figure 2 and its animation). In the bottom panels, a red tick marks the delta sunspot's PIL centered on the downflow. The black bar spans the delta sunspot's nominal diameter.

Figure 6b. Schematic of Scenario I-B for Type I genesis of the sharp-PIL delta sunspot in AR 12876 from a nearly east-west 180° writhe kink in a horizontal flux rope that was directed east-west when it kinked. The vertical plane, symbolism, and color scheme are the same as in Figure 6a. The flux rope's field has right-handed twist, the kink's positive-polarity leg is against the plane's near side, and the kink's negative-polarity leg is against the plane's far side. During its emergence and partial submergence, the kink is horizontally trapped in downflow between convection cells. Because the flux-rope field's twist is right-handed and its twist pitch decreases inward to zero at the center field line, the kink's BMR footprint pivots counterclockwise in photospheric magnetograms and continuum images as the kink emerges and the kink's coronal field becomes an S-shaped sigmoid, as observed for AR 12876 (Figure 2 and its animation). In the bottom panels, a red tick marks the delta sunspot's PIL centered on the downflow. The black bar spans the delta sunspot's nominal diameter.

Figure 7. Schematic for our scenario for Type II genesis of a sharp-PIL delta sunspot as in AR 12175, but for head-to-tail collision of two same-size emerging east-west $\Omega$ loops centered on the same east-west vertical plane. The symbolism and color scheme are the same as in Figure 6a. Each $\Omega$ loop emerges at the same time in the upflow of one of a pair of adjacent same-size convection cells. For each $\Omega$ loop, the east leg is positive and the west leg is negative. As the loops emerge, the western loop's positive flux and the eastern loop's negative flux are swept into the central downflow while each outer foot is swept into the nearest outer downflow. The central convergence of opposite-polarity flux makes a sharp-PIL delta sunspot centered on the central downflow. In the bottom panel, the sharp PIL is viewed end-on at the red tick mark, and the black bar spans the delta-sunspot's nominal diameter.

Figure 8. Schematic of our scenario for Type III genesis of a sharp-PIL delta sunspot similar to the genesis in AR 13190, but for tail-on emergence of a small BMR against an edge of a big sunspot. As in our other schematics, the plane of the drawings is vertical, east-west, viewed from the south, bisects lengthwise a rising east-west $\Omega$ loop, and, here, bisects a big sunspot just east of the $\Omega$ loop. Otherwise, the symbolism and color scheme are the same as in Figure 6a. <u>Top</u>: The $\Omega$ loop is rising in the upflow of a convection cell at edge of the big negative sunspot and is about to emerge under the sunspot's magnetic canopy. <u>Middle</u>: Convection sweeps the emerging $\Omega$ loop's positive leg against the negative canopy field, driving reconnection with the canopy and making a very lopsided delta sunspot having a sharp PIL (red tick) enveloped by a magnetic arcade. The $\Omega$ loop's western leg is being swept into the convection cell's western downflow, and that downflow is migrating westward. <u>Bottom</u>: The $\Omega$ loop continues emerging and reconnecting with the canopy to further build the magnetic arcade over the delta sunspot's sharp PIL. The convection cell's western downflow has migrated farther west and the $\Omega$ loop's western (negative) leg is now in the western downflow.  <u>Middle and Bottom</u>: The black bar spans the western half of the very lopsided delta sunspot.

Figure 9. Schematic of our scenario for Type IV genesis of a sharp-PIL delta sunspot similar to the genesis in AR 13014, but for head-to-tail merging of two east-west emerging Ω loops that are centered on the same east-west vertical plane.  The symbolism and color scheme are the same as in Figure 6a.  <u>Top</u>: The two Ω loops are rising in adjacent convection cells of unequal size and are on the verge of emerging. Each leg of each Ω loop is being swept into the downflow nearest that leg.  <u>Middle</u>: Both Ω loops are emerging.  The eastern-most downflow in the top panel has died out and has been replaced by horizontal westward convection into the downflow at the eastern edge of western Ω loop's convection cell.  The eastern Ω loop's leading (positive) leg and the western Ω loop's trailing (negative) leg are being swept into that downflow. The western Ω loop's leading (positive) leg is being swept into the western downflow.  <u>Bottom</u>: Both legs of the eastern Ω loop and the negative leg of the western Ω loop have been crammed together into the eastern downflow.  That has made a delta sunspot in which there are two sharp PILs (red ticks).  The eastern Ω loop envelops the eastern sharp PIL, and reconnection of the western Ω loop's negative leg with the eastern Ω loop's positive leg has made a magnetic arcade enveloping the western sharp PIL.  The black bar span's the delta sunspot's nominal diameter.

| Genesis Type | Definition and Schematic |
|---|---|
| Type I | Inside merger of a single BMR's opposite-polarity flux. 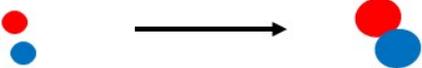 |
| Type II | Outside merger of leading end of an emerging east-west BMR with opposite-polarity trailing end of another emerging east-west BMR. 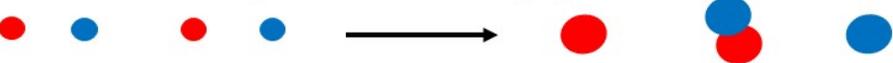 |
| Type III | Emergence of a relatively small BMR at an edge of a unipolar larger sunspot. 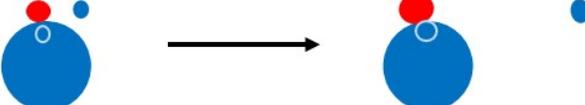 |
| Type IV | Any genesis that is not Type I, II, or III, such as merger of an emerging BMR with itself and with another emerging BMR, or merger of two emerging BMRs with themselves and with larger unipolar flux. 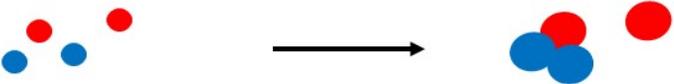 |

**Figure 1**

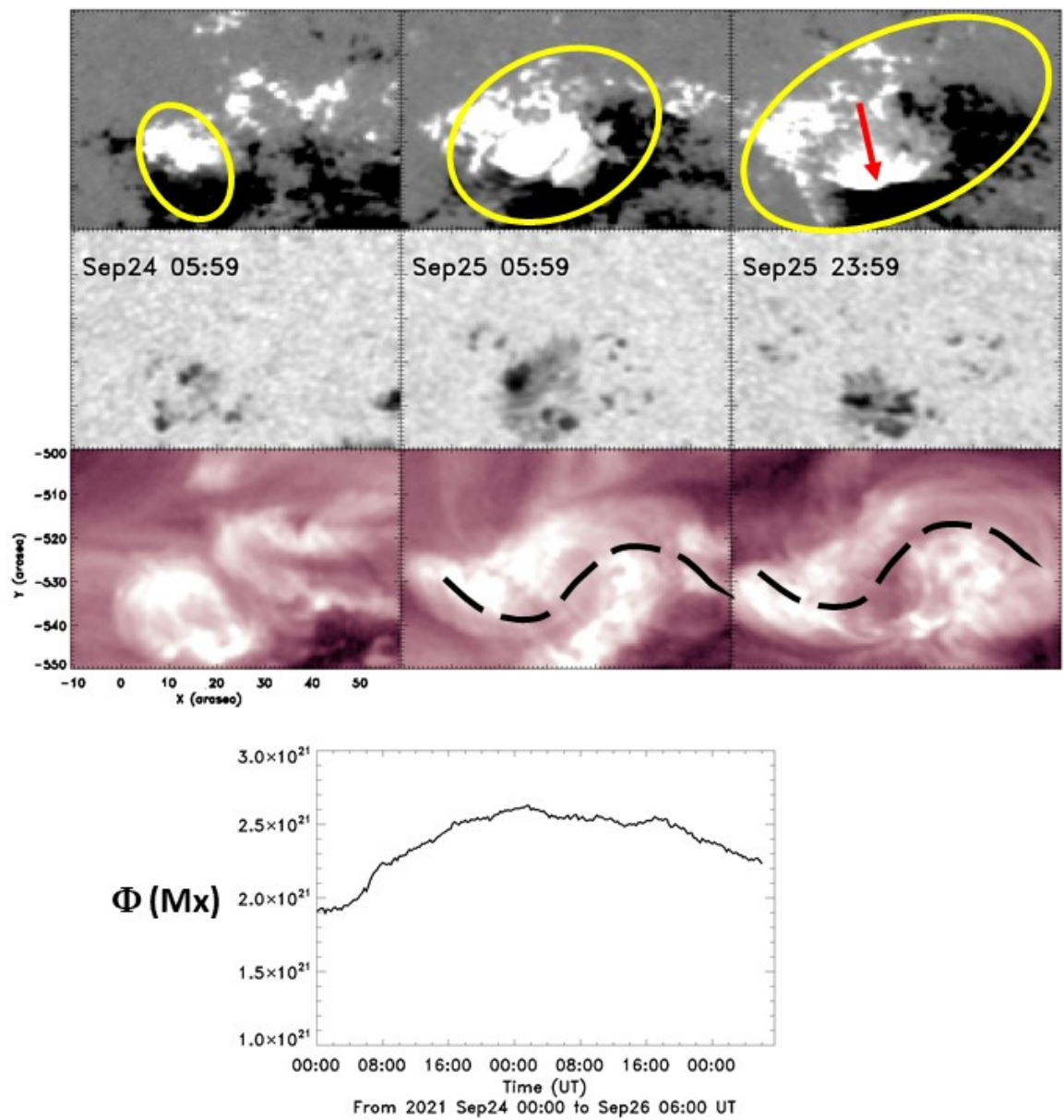

**Figure 2**

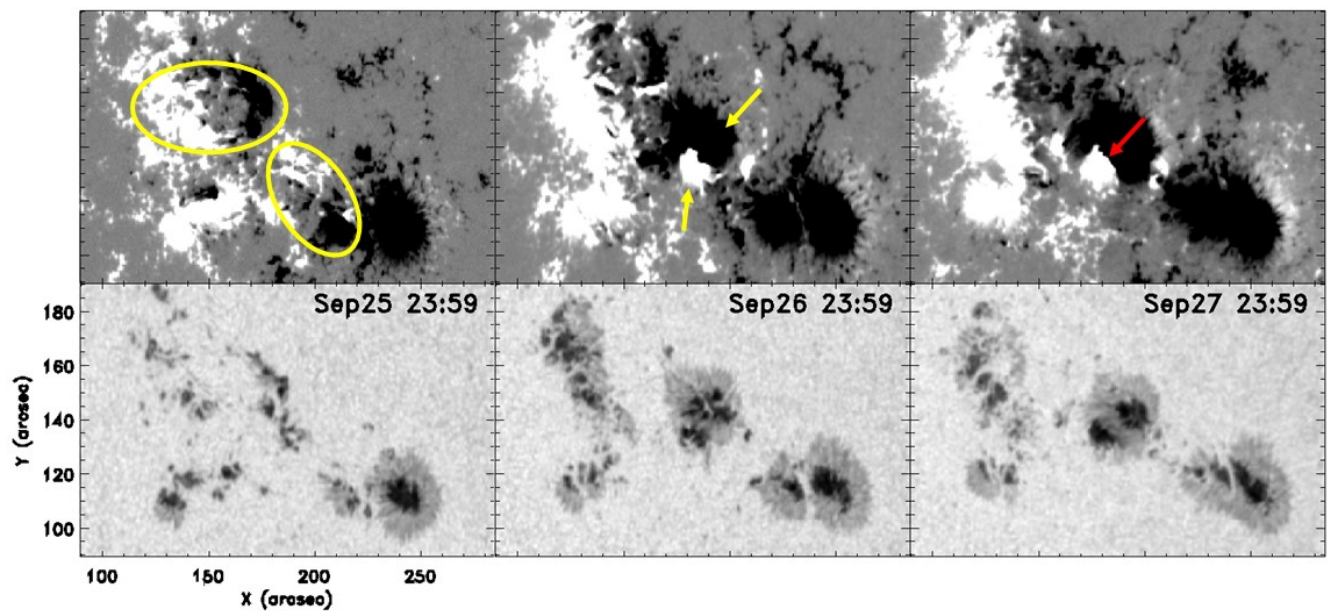

Figure 3

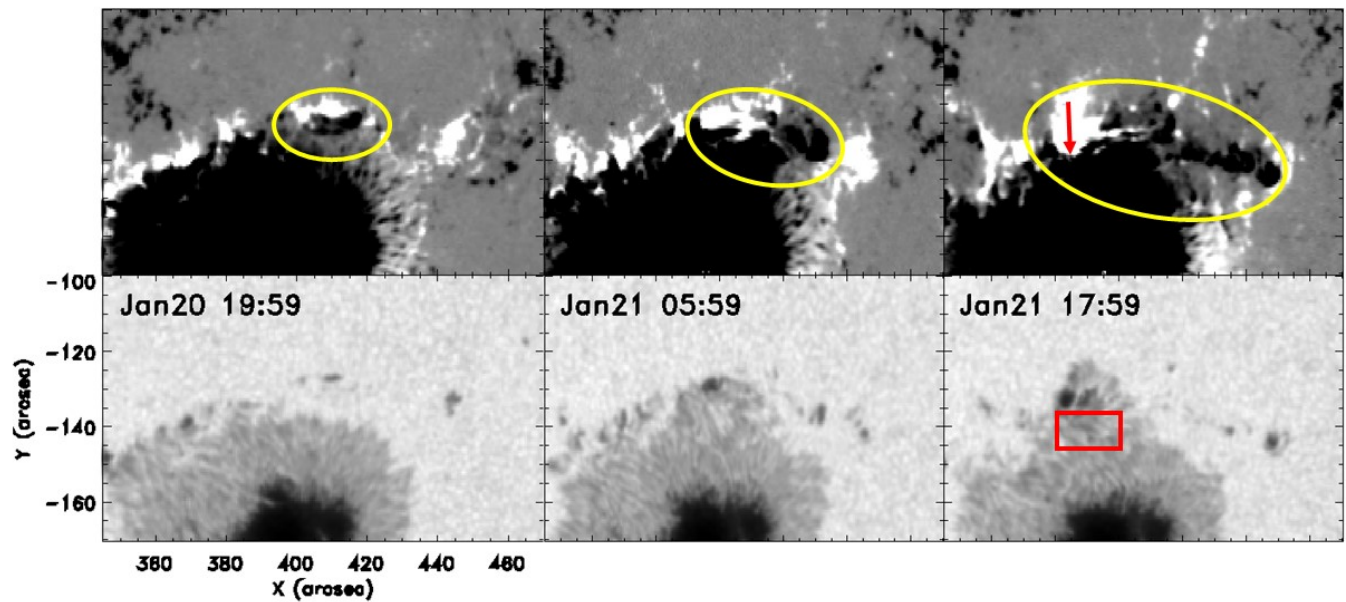

Figure 4

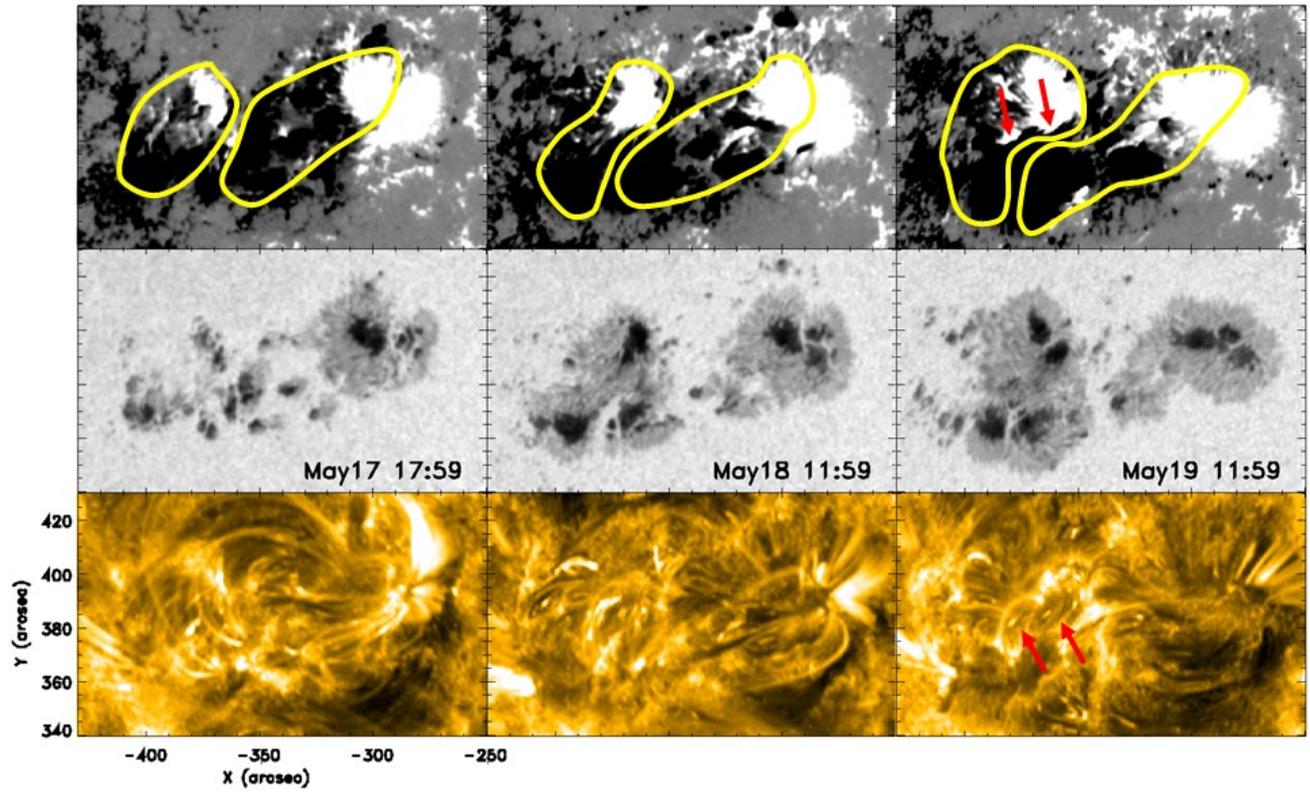

**Figure 5**

## Emergence Impending

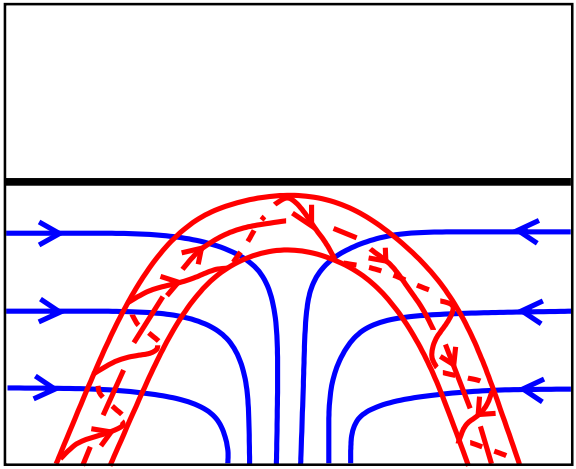

## Early in Emergence

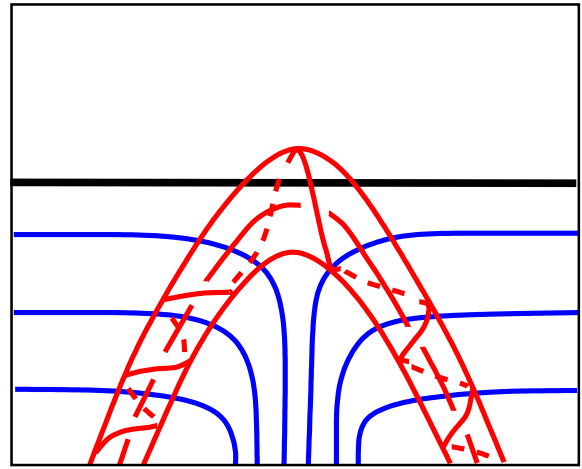

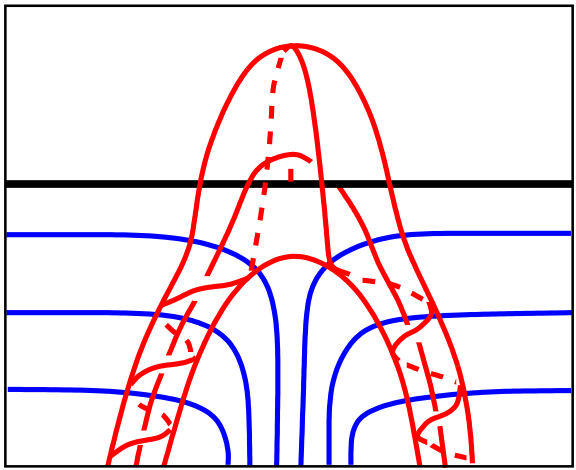

**Emergence Ending**

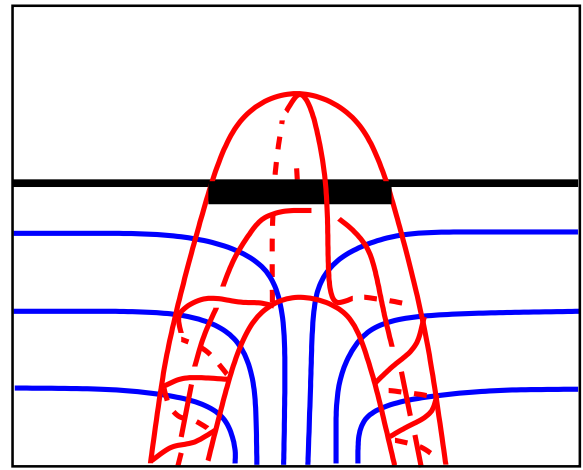

**Mid Submergence**

**Figure 6a**

## Emergence Impending

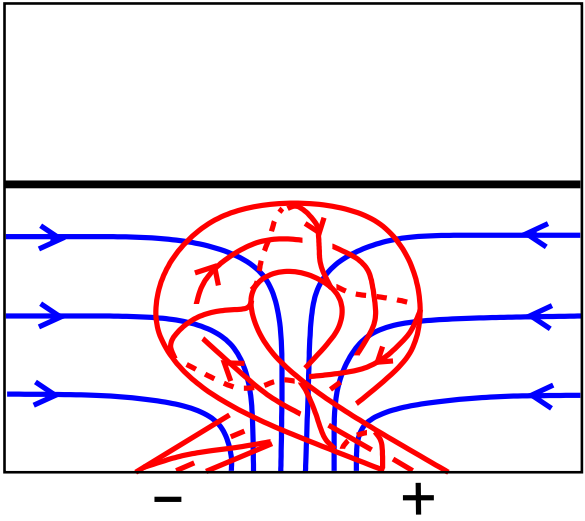

## Early in Emergence

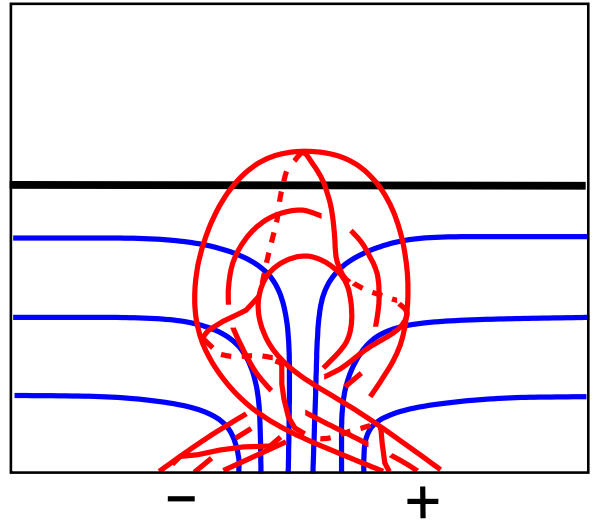

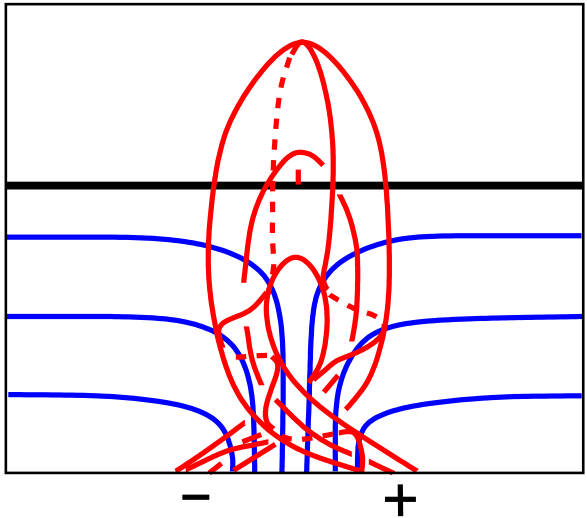

## Emergence Ending

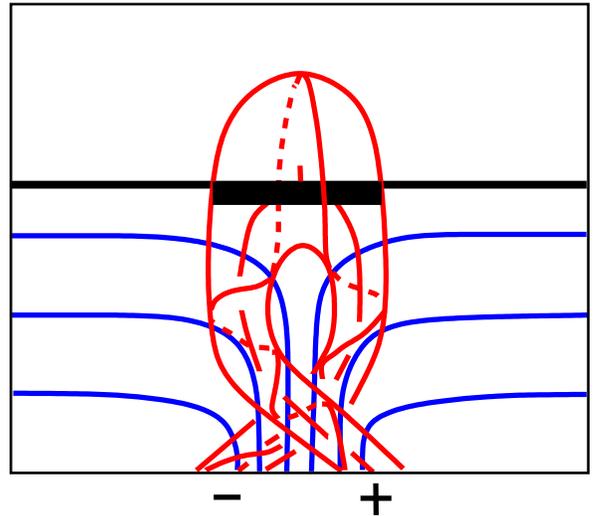

## Mid Submergence

**Figure 6b**

## Impending Emergence of Two BMRs, Head−to−Tail

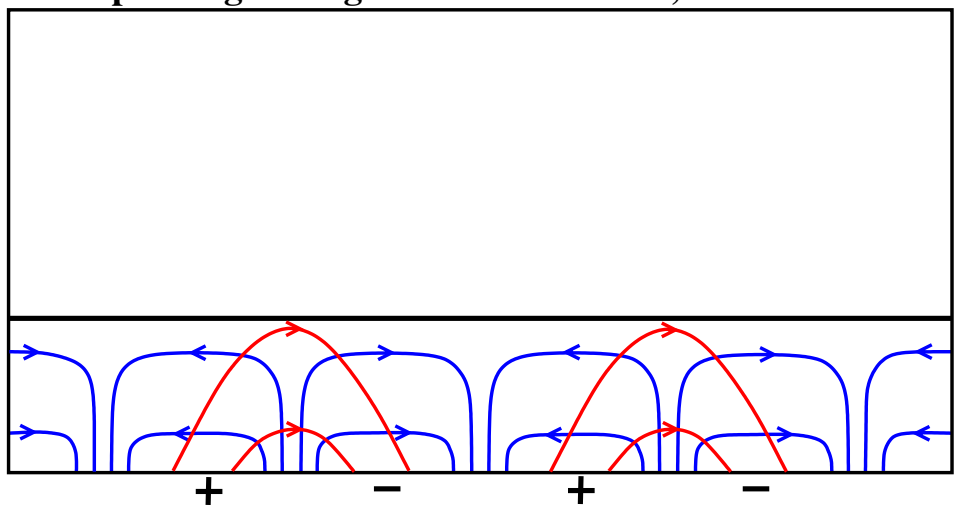

## Emergence Leading to Sharp PIL

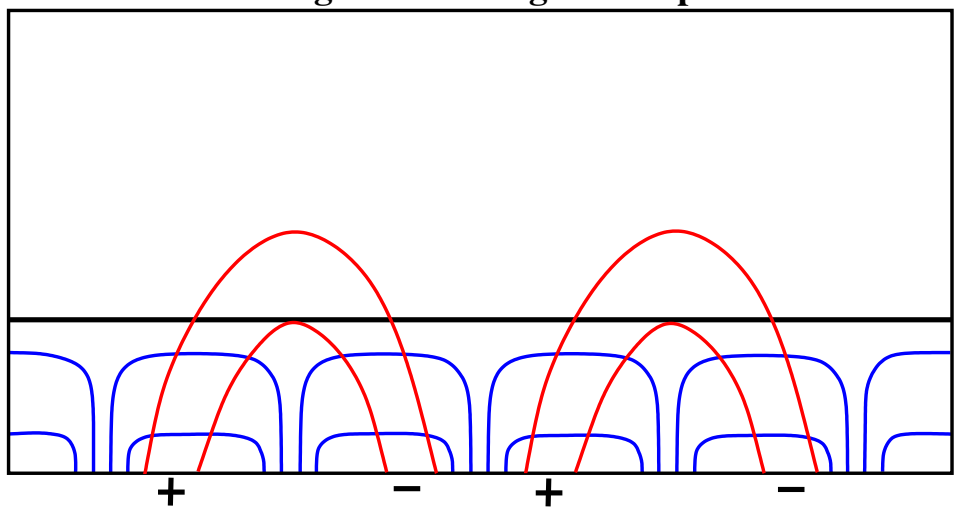

## Sharp−PIL Genesis Ending

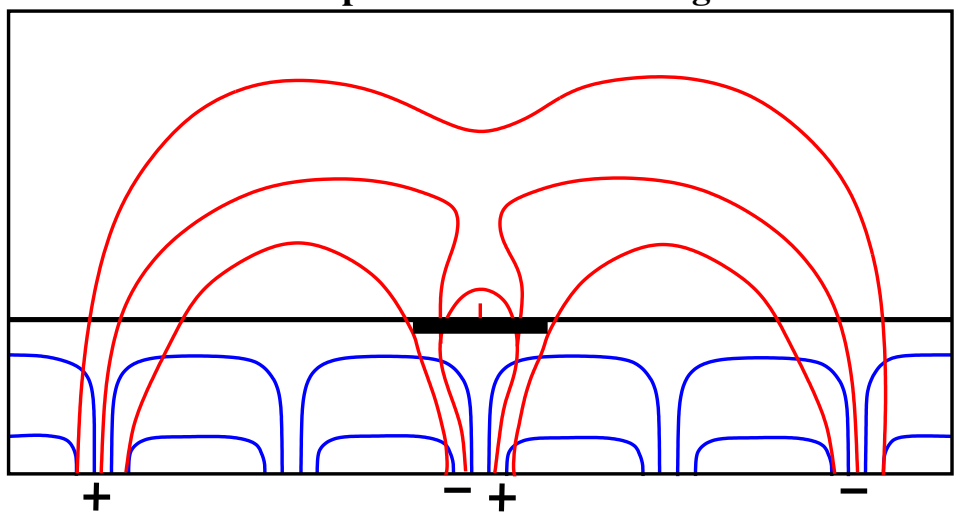

**Figure 7**

**Impending Emergence of Small BMR at Edge of Much Bigger Sunspot**

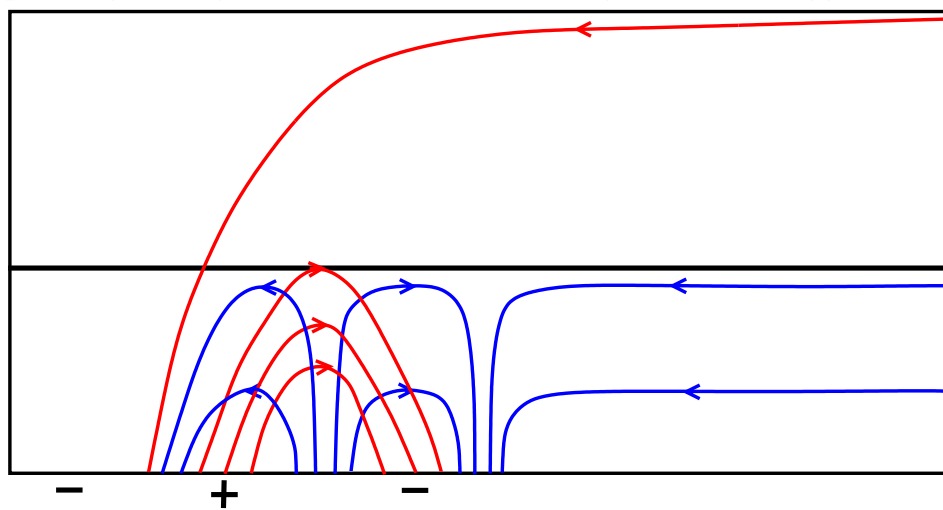

**Emergence Leading to Sharp PIL**

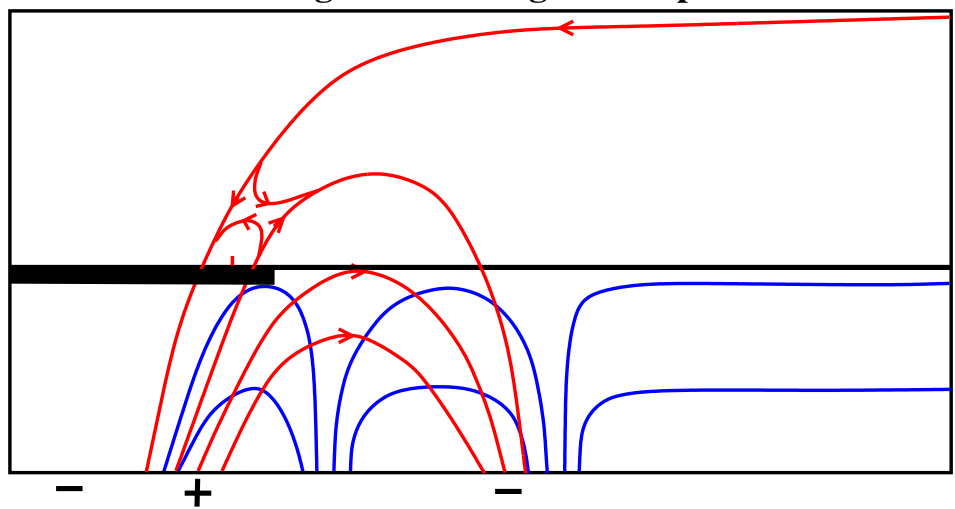

**Sharp−PIL Genesis Ending**

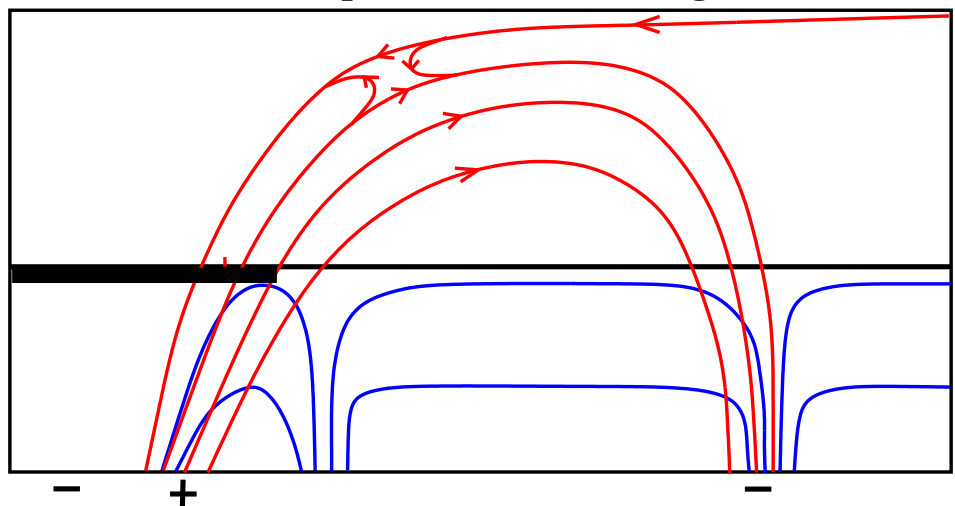

**Figure 8**

## Impending Emergence of Two BMRs, Head−to−Tail

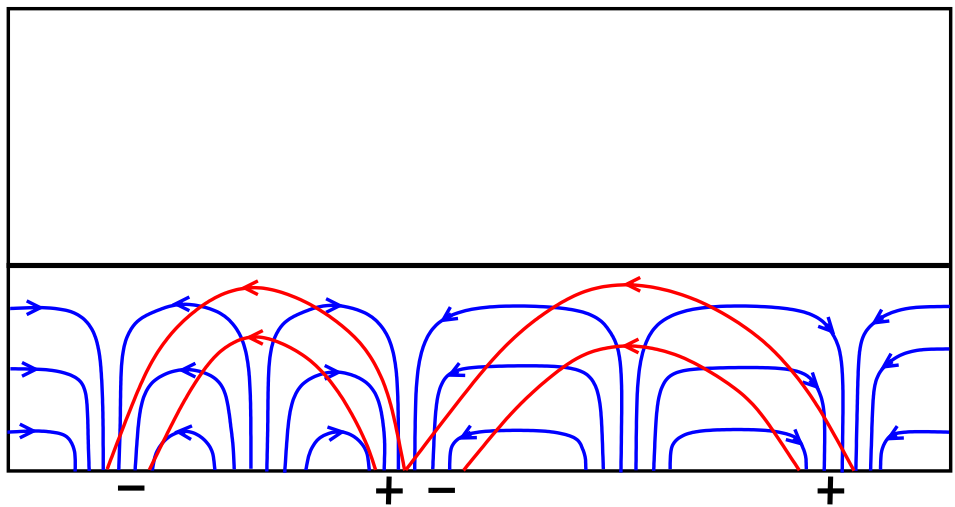

## Emergence Leading to Sharp PIL

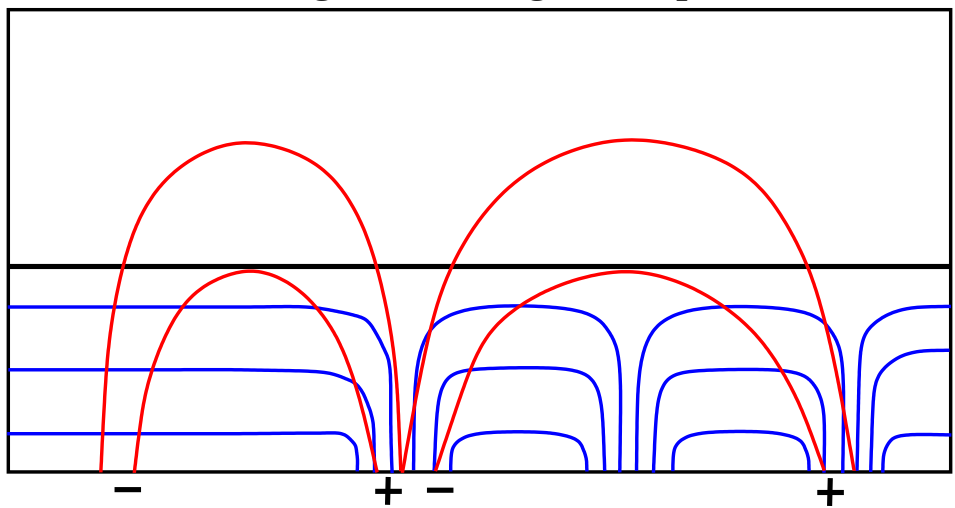

## Sharp−PIL Genesis Ending

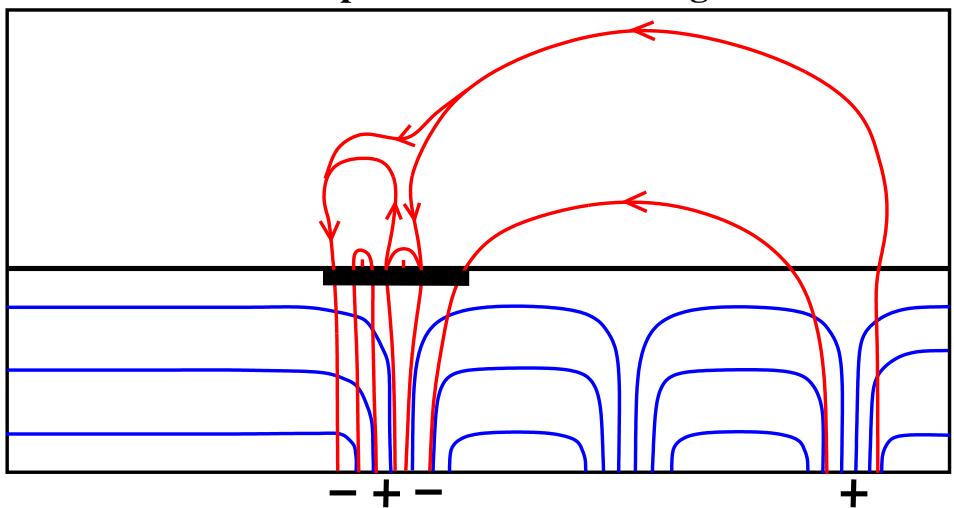

**Figure 9**